\documentclass[noacm, sigconf]{acmart}

\makeatletter
\def\@ACM@copyright@check@cc{}
\makeatother

\copyrightyear{2025}
\acmYear{2025}
\setcopyright{cc}
\setcctype{by}
\acmConference[SC '25]{The International Conference for High Performance Computing, Networking, Storage and Analysis}{November 16--21, 2025}{St Louis, MO, USA}
\acmBooktitle{The International Conference for High Performance Computing, Networking, Storage and Analysis (SC '25), November 16--21, 2025, St Louis, MO, USA}
\acmDOI{10.1145/3712285.3759816}
\acmISBN{979-8-4007-1466-5/2025/11}

\AtBeginDocument{%
  }

\acmSubmissionID{pap397}

\usepackage{xspace}
\usepackage{tabularx}
\usepackage{makecell}

\definecolor{dong}{RGB}{0,0,0}
\definecolor{check}{RGB}{0,0,0}
\definecolor{sherry}{RGB}{255,128,0}
\definecolor{draft}{RGB}{102,204,0}
\definecolor{bin}{RGB}{100, 000, 200}

\begin{document}

\newcommand{\name}{cMPI\xspace}
\title{\name: Using CXL Memory Sharing for MPI One-Sided and Two-Sided Inter-Node Communications}

\author{Xi Wang}
\authornote{This work was done during the first author's internship at SK hynix.}
\orcid{0009-0001-6251-8177}
\affiliation{%
  \institution{University of California, Merced}
  \city{Merced}
  \state{CA}
  \country{USA}
}
\email{swang166@ucmerced.edu}

\author{Bin Ma}
\affiliation{%
  \institution{University of California, Merced}
  \city{Merced}
  \state{CA}
  \country{USA}
}
\email{bma100@ucmerced.edu}

\author{Jongryool Kim}
\affiliation{%
  \institution{SK hynix}
  \city{San Jose}
  \state{CA}
  \country{USA}
}
\email{jongryool.kim@sk.com}

\author{Byungil Koh}
\affiliation{%
  \institution{SK hynix}
  \city{San Jose}
  \state{CA}
  \country{USA}
}
\email{byungil.koh@sk.com}

\author{Hoshik Kim}
\affiliation{%
  \institution{SK hynix}
  \city{San Jose}
  \state{CA}
  \country{USA}
}
\email{hoshik.kim@sk.com}

\author{Dong Li}
\affiliation{%
  \institution{University of California, Merced}
  \city{Merced}
  \state{CA}
  \country{USA}
}
\email{dli35@ucmerced.edu}

\renewcommand{\shortauthors}{Wang et al.}

\begin{abstract}

Message Passing Interface (MPI) is a foundational programming model for high-performance computing. 
\textcolor{check}{MPI libraries traditionally employ network interconnects (e.g., Ethernet and InfiniBand) and network protocols (e.g., TCP and RoCE) with complex software stacks for cross-node communication.}
\textcolor{check}{This paper presents \name\footnote{\name\ is open-sourced at \url{https://github.com/skhynix/cMPI}.}, the first work to optimize MPI point-to-point communication (both one-sided and two-sided) using CXL memory sharing on a real CXL platform, transforming cross-node communication into memory transactions and data copies within CXL memory, bypassing traditional network protocols. }
\textcolor{check}{We analyze performance across various interconnects and find that CXL memory sharing achieves 7.2$\times$-8.1$\times$ lower latency than TCP-based interconnects deployed in small- and medium-scale clusters. } 
\textcolor{check}{We address challenges of CXL memory sharing for MPI communication, including} data object management over the \texttt{dax} representation~\cite{ma2024hydrarpc}, cache coherence, and atomic operations. 
\textcolor{check}{Overall, \name outperforms TCP over standard Ethernet NIC and high-end SmartNIC by up to 49$\times$ and 72$\times$ in latency and bandwidth, respectively, for small messages.}

\end{abstract}



\maketitle

\section{Introduction}
\label{sec:intro}
Message Passing Interface (MPI) is the de-facto standard to program distributed memory in high-performance computing (HPC), and has been utilized to enable large-scale process-level parallelism for many scientific and engineering applications. Based upon data copy and message passing, MPI allows processes with separate address spaces to communicate and share data. 


Given two MPI processes distributed on two nodes, \textcolor{check}{two types of MPI communication can be employed}, either one-sided or two-sided depending on process participation. 
The MPI communication traditionally employs high-performance interconnect (e.g., Ethernet), and relies on Remote Direct Memory Access (RDMA) or network protocols (e.g.,  RoCE~\cite{10.1145/3543176}). The support of high-performance interconnect includes a complicated software stack, including application-level protocol implementation and system-level driver for device registration and interrupt handling. Using the interconnect technologies also requires the MPI library to use special APIs or programming methods  (e.g., socket programming).

Recent work~\cite{10.1145/3488423.3519336,10.1145/3581784.3607108,10.1007/978-3-031-32041-5_16} reveals low memory utilization in individual nodes in data centers or (supercomputers), and demonstrates the potential of memory disaggregation for improving memory utilization and reducing production cost on  supercomputers~\cite{10.1145/3624062.3624175,10.1145/3581784.3607108,10.1145/3575693.3578835}. 
The emerging Compute Express Link (CXL)~\cite{sharma2023introduction} makes memory disaggregation highly feasible. CXL is an interconnect technology designed for high-speed communication among processors, accelerators, and memory devices.  
CXL allows for memory expansion beyond traditional CPU memory channels by enabling a CPU to communicate with DRAM managed by its own controller over a PCIe channel, hence offering increased capacity and bandwidth without increasing the number of primary CPU memory channels. The CXL memory can be accessed using traditional CPU instructions (e.g., load and store), allowing the existing applications to use it with little changes. The CXL memory has been explored as a memory tier to build HPC systems with larger memory capacity for checkpointing or accommodating scientific applications with larger memory footprint~\cite{10024061,10.1145/3624062.3624175,10898637,wang2025cxl_performance,10.1145/3581784.3607108}. 

\begin{figure}[tb!]
	\centering
         \includegraphics[width=0.9\linewidth]{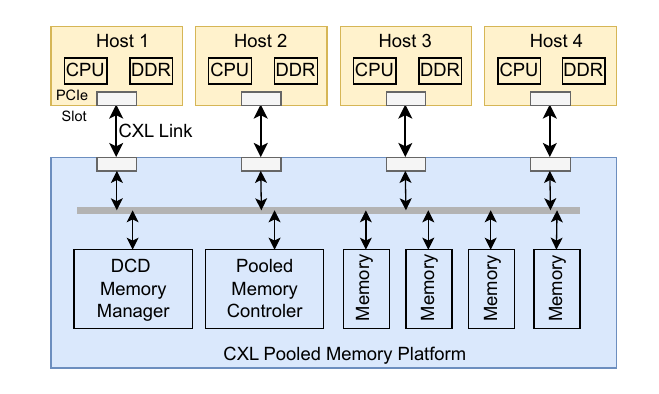}
	\caption{CXL memory sharing using a CXL pooled-memory platform. The CXL memory sharing is based upon the Multi-Headed Device (MHD) and Dynamic Capacity Device (DCD). }
	\centering
	\label{fig:CXL_MHD} 
\end{figure}

In this paper, we study a new MPI  communication scheme using CXL memory sharing. The CXL memory sharing is based upon the CXL pooled memory platform (a solution for memory disaggregation), shown in Figure~\ref{fig:CXL_MHD}. Using the CXL memory sharing, the MPI communication across nodes can be transformed into regular memory transactions and data copy within the CXL memory, without going through the traditional network protocols. There are a couple of benefits of doing so.
First, the communication latency can be  shorter, compared with network‑based approaches such as TCP over Ethernet and RoCEv2.
For example, given a message size of 8 bytes, we show that using a \textit{real} CXL pooled-memory platform (not based on hardware simulation or emulation~\cite{ahn2022enabling,memsys_cxl_switch_2024,osu_omb}) to access the CXL memory takes only 790 $ns$, while using TCP over a standard Ethernet NIC and TCP over Mellanox (CX-6 Dx) to access data across nodes take 16 $\mu s$ and 18 $\mu s$ respectively, about an order of magnitude longer than using the CXL memory sharing (see Section~\ref{sec:bg} for details). 
Second, using the CXL memory sharing leads to a simpler software stack, compared to using the traditional network protocols. This is because the CXL protocol to perform CXL memory transactions for data (de)packaging and synchronization between CPU and memory device are done by hardware (not system software). 


However, using the CXL memory sharing for MPI communication across nodes faces challenges. First, the representation of the CXL memory sharing in the host (node) creates challenges to manage data objects in the CXL memory. At the node, the CXL memory sharing is exposed as a Direct Access (\texttt{dax}) device by the CXL driver and {\texttt{daxctl}~\cite{ma2024hydrarpc}. The \texttt{dax} device bypasses the traditional page cache and provides memory-level access. However, the abstract of the \texttt{dax} device cannot provide flexibility to create arbitrary data objects and manage their life cycle (e.g., creation and deletion). 


Second, the CXL hardware creates difficulty for the MPI library to implement one-sided and two-sided communications because of cache coherence and atomic operations. For high-performance communication, the MPI library often uses an internal buffer to hold messages and enable asynchronous operations. This buffer, when allocated on the CXL memory sharing, must be cache-coherent to ensure execution correctness. The CXL protocol defines hardware-based cache coherence. However, such a cache coherence mechanism is not scalable because of tracking cache line statuses at a large scale and potentially with large CXL memory. Furthermore, the MPI library often uses atomic operations to coordinate concurrent accesses to internal message queues. However, the CXL pooled memory often lacks a mechanism to enforce atomicity across nodes.


To address the above challenges and make feasible the CXL memory sharing for MPI one-sided and two-sided communications across nodes, we introduce \textit{\name}. \name is an extension to the MPI library and does not require any change to the application and other system software. To address the first challenge, \name introduces a data-object management system without changing the \texttt{dax} abstract for the CXL memory sharing. By mapping the CXL memory into the virtual address space of MPI process and bookkeeping the information of data objects based on multi-level hashing, \name brings flexibility to create arbitrary data objects. To address the second challenge, \name introduces software-based cache coherence and enforces it on demand only between communicating processes. This solution is lightweight and does not require hardware changes. To tackle the problem of lacking atomicity for managing message queues, we propose a message queue structure for pair-wise MPI communication across nodes, and hence avoid the needs of atomic operations. Built upon the above design, \name further introduces techniques to support message management and synchronization.  

Our contributions are summarized as follows.

\begin{itemize}
    \item We explore the feasibility and benefits of using the CXL memory sharing for MPI one-sided and two-sided  communications across nodes on real CXL memory. \textcolor{check}{Our analysis shows that while high-end RDMA-supported devices deliver superior performance, CXL memory sharing shows significant performance potential compared to TCP-based interconnects commonly deployed in small- and medium-scale clusters.}

    \item We develop a library, \name, by extending MPICH-4.2.3~\cite{MPICH_online}. We address the limitation of CXL memory sharing to support MPI communications. 
   
    \item Our evaluation shows that using the CXL memory sharing for one-sided and two-sided inter-node communications, \name outperforms TCP over a standard Ethernet NIC by up to 49$\times$ in latency and 72$\times$ in bandwidth for all messages, and outperforms TCP over a high-end SmartNIC by up to 48$\times$ in latency and 3.7$\times$ in bandwidth for small messages.
\end{itemize}
 
\section{Background and Motivation}
\label{sec:bg}
We review the background and motivate our work in this section.

\subsection{MPI}

MPI provides a set of routines offering a portable way to express parallel communication patterns. 
For inter-node communication (i.e., the communication between processes on different hosts), MPI utilizes network stacks to transfer messages. For intra-node communication (i.e., the communication between processes on the same host), MPI utilizes shared memory to transfer messages. 
We use the term ``MPI process'' and ``rank'' interchangeably in the paper.

\textbf{One-sided communication} 
is built upon the scheme of Remote Memory Access (RMA). It utilizes an RMA window, 
which allows an MPI process to directly access data residing in another MPI processes. The process initiating data access is called the origin process, while the process whose data is accessed is called the target process. Unlike the two-sided communication, the one-sided communication requires only the origin process to execute RMA operations such as \texttt{MPI\_Get} (fetching data from a target process)  and \texttt{MPI\_Put} (storing data into a target process). Because of the absence of the coordination between the origin and target processes, there is a need for synchronization between the two processes.

\textbf{Two-sided communication} explicitly involves a sender and a receiver during the communication. 
It utilizes intermediate buffers (MPI internal buffer and NIC buffer) to transfer data between two MPI processes since a sender process cannot directly access a receiver process's address space. This pattern requires both the sender and receiver participate in the communication by calling matching \texttt{MPI\_Send} and \texttt{MPI\_Recv} routines (or their non-blocking version). 

Figure~\ref{fig:MPI_communication} shows the overall workflow of traditional MPI one-sided and two-sided inter-node communications. 

\begin{figure}[tb!]
	\centering
         \hbox{\hspace{-1em}\includegraphics[width=1.1\linewidth]{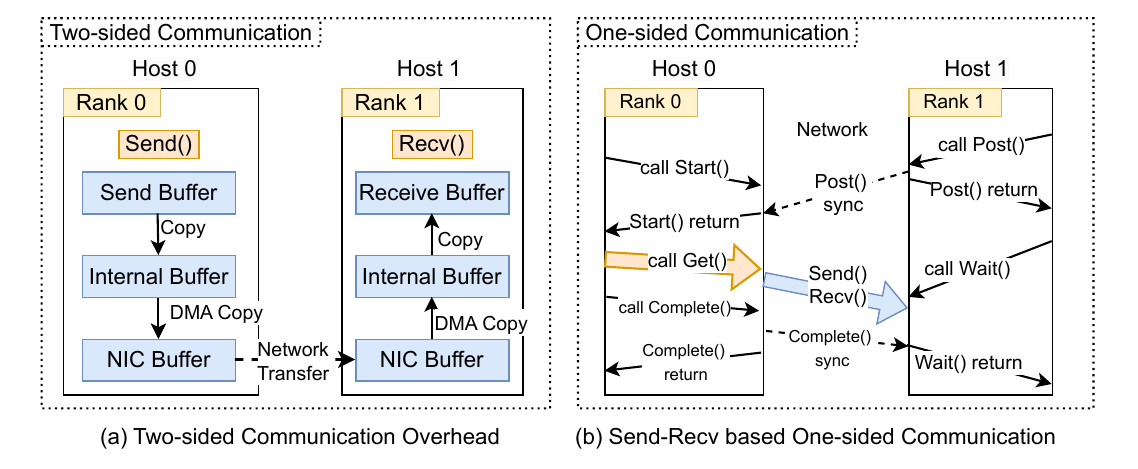}}
	\caption{MPI communication across hosts (nodes).}
	\centering
	\label{fig:MPI_communication} 
\end{figure}

\subsection{CXL}

\begin{table}[!t]
    \centering
    \caption{Memory access latency and bandwidth over various interconnect and protocols.}
    \resizebox{\linewidth}{!}{%
    \begin{tabular}{|c|c|c|}
        \hline
        \textbf{Arch Type} & \textbf{Latency} & \textbf{Bandwidth} \\
        \hline
        \addlinespace[4pt]
        \hline
        \texttt{Main Memory} & 100 $ns$ & 132.8 GB/s \\
        \hline
        \makecell[c]{\texttt{TCP over Standard Ethernet NIC}\\\texttt{(abbreviated as: TCP over Ethernet)}} 
            & 16 $\mu s$ & 117.8 MB/s \\
        \hline
        \makecell[c]{\texttt{TCP over Mellanox (CX-6 Dx)}} 
            & 18 $\mu s$ & 11.5 GB/s \\
        \hline
        \makecell[c]{\texttt{RoCEv2 over Mellanox (CX-6 Dx)}} 
            & 1.6 $\mu s$ & 10.8 GB/s \\
        \hline
        \makecell[c]{\texttt{RoCEv2 over Mellanox (CX-3)}} 
            & sub-2 $\mu s$ & 7.0 GB/s \\
         \hline
        \makecell[c]{\texttt{InfiniBand over Mellanox (CX-6)}} 
            &  sub-600 $ns$ & 25.0 GB/s \\
         \hline
        \makecell[c]{\texttt{CXL Memory Sharing}\\\texttt{\small (with caching; no cache flushing)}} & 790 $ns$ & 9.9 GB/s \\
        \hline
        \makecell[c]{\texttt{CXL Memory Sharing}\\\texttt{(with cache flushing)}} & 2.2 $\mu s$ & 9.5 GB/s \\
        \hline
    \end{tabular}
    }
    \label{tab:memory_type}
\end{table}

The recent CXL specification (CXL 3.0) introduces CXL memory sharing. Based upon the Multi-Headed Device (MHD) and Dynamic Capacity Device (DCD), a host (node) can connect to the CXL pooled memory platform through a dedicated CXL port, ensuring bandwidth fairness and eliminating added latency of a switch hop, shown in Figure~\ref{fig:CXL_MHD}. We use a CXL pooled memory platform, Niagara 2.0 \cite{Niagara20}, for CXL memory sharing.


\textbf{CXL memory sharing} allows multiple processors to access a shared memory region, offering a unique opportunity for efficient communication between processors. First, data can be exchanged between processors without relying on additional communication protocols such as TCP/IP or RoCE. Second, the programming model is simplified, as each processor interacts with the shared memory using CPU instructions (e.g., load and store), mirroring the behavior of local memory accesses. Finally, the memory sharing through CXL can offer lower communication latency, compared to traditional network-based communication mechanisms. 

\textbf{Performance evaluation.} To assess the benefits of using CXL memory sharing to improve  communication performance across nodes, we compare the following 8 cases.  
\begin{enumerate}
    \item The traditional main memory attached to the CPU through memory bus (or ``main memory'' for short);
    \item TCP over a standard Ethernet NIC (or ``\textit{TCP over Ethernet}'' for short);
    \item TCP over Mellanox CX-6 Dx (a high-end SmartNIC); 
    \item RDMA over Converged Ethernet v2 (i.e., RoCEv2) over Mellanox CX-6 Dx; 
    \item RoCEv2 over Mellanox CX-3 (a low-end SmartNIC);
    \item InfiniBand over Mellanox CX-6;
    \item CXL memory sharing (with caching, but without using cache flushing for cache coherence);
    \item CXL memory sharing with cache flushing for cache coherence (see Section~\ref{sec:design:consistency}). 
\end{enumerate}

Section~\ref{sec:eval:method} has more hardware details. The case (1) studies intra-node communication, providing the best performance. The other cases study inter-node communication on two nodes. 


To measure latency and bandwidth, we use Intel MLC \cite{mlc} for the cases (1) and (7), iPerf~\cite{iPerf_online} for (2) and (3), and Perftest~\cite{perftest} for (4). We choose those different tools (benchmarks) based on whether they are aligned with the specific system features (e.g., RDMA, or  TCP over Ethernet). As we do not have Mellanox CX-3 and CX-6, the performances for the cases (5) and (6) are taken from the vendor's product report \cite{MellanoxCX3,MellanoxCX6}. For the case (8), 
since existing tools do not support testing memory access with cache flushing on a \texttt{dax} device (the system-level abstraction for CXL memory sharing), we develop a micro-benchmark. This benchmark uses \texttt{mmap} to obtain the virtual address space of the \texttt{dax} device. For latency tests, this benchmark performs \texttt{memset} with cache flushing. During the test, we vary the data size and measure the average latency over 1000 iterations. For bandwidth tests, this benchmark uses multiple threads, and each thread accesses 512 bytes with cache flushing on a \texttt{dax} device region. We measures aggregated bandwidth. See Table \ref{tab:memory_type} for results. 


\textbf{Observation 1.} The CXL memory sharing (with cache flushing) outperforms TCP over Ethernet and TCP over high-end SmartNIC. 

The latency of CXL memory sharing is 7.2$\times$-8.1$\times$ lower \textcolor{check}{than that of TCP-based interconnect}. Compared with TCP over Ethernet, the bandwidth of CXL memory sharing is 80$\times$ higher. The performance benefit of the CXL memory sharing is because it allows direct load/store operations to remote memory, eliminating the need to wrap data in network packets, traverse the full TCP/IP stack, and perform additional context switches. 

\textbf{Observation 2.} The CXL memory sharing (with cache flushing) has longer latency but comparable bandwidth than RoCEv2, and performs worse than InfiniBand (on Mellanox CX-6). 

\textbf{Observation 3.} Cache flushing in the CXL memory sharing for cache coherence increases latency by 2.8$\times$.


In conclusion, while using high-end RDMA-supported devices for inter-node communication delivers the best performance, using the CXL memory sharing for inter-node communication shows great performance potential, compared with TCP-based interconnect which is commonly employed in small- or middle-scale clusters. Furthermore, if the CXL memory sharing can reduce the latency of cache flushing, the latency of using the CXL memory sharing for inter-node communication can be more competitive. We refer to the CXL memory sharing as \textit{CXL SHM} in the remaining sections.


\section{Design}
\label{sec:design}


The goals of \name are to enable MPI processes to freely create shareable memory objects for message passing and to significantly reduce communication overhead across nodes. \name does not modify the user-facing MPI routines (e.g., MPI initialization, message-passing, and synchronization interfaces). 
Figure~\ref{fig:overview} overviews \name.  Multiple nodes connect to a CXL pooled memory platform through CXL physical links, sharing data stored in the CXL memory. We design a management layer called the \textit{CXL SHM Arena} to efficiently allocate independent shared memory objects. The nodes involved in MPI communication exchange data by storing and loading directly from the CXL shared memory. For the two-sided communication, the message queues used in the traditional MPI library are created and maintained in the CXL shared memory, allowing nodes to transfer messages via enqueue and dequeue operations on the CXL shared memory. For the one-sided communication, Remote Memory Access (RMA) windows are allocated within the CXL shared memory, allowing processes to execute RMA operations such as \texttt{MPI\_Get()} and \texttt{MPI\_Put()} by directly accessing remote ranks' data stored in the CXL shared memory, eliminating network-based data transfers.

\begin{figure}[tb!]
	\centering
         \includegraphics[width=1.0\linewidth]{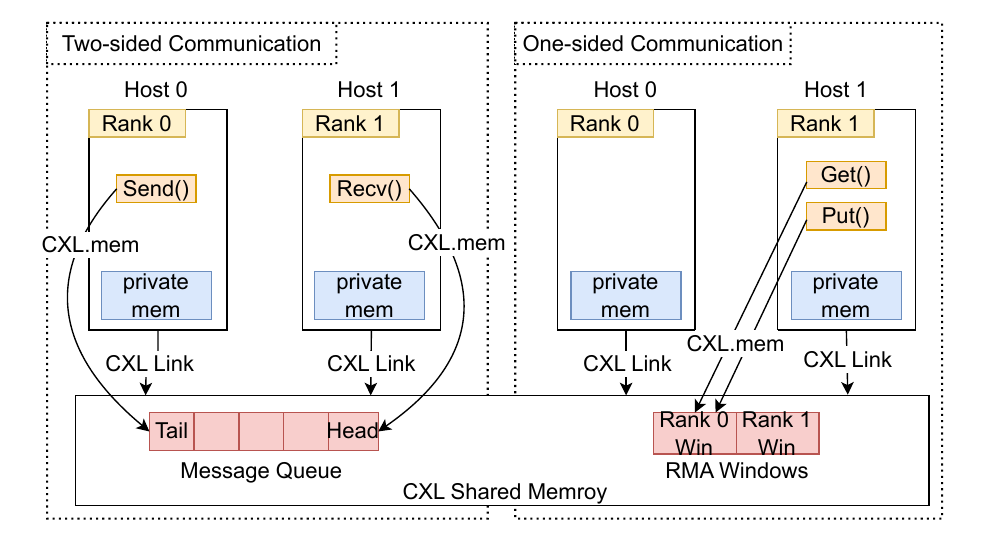}
	\caption{The overview of \name.}
	\centering
	\label{fig:overview} 
\end{figure}


\subsection{CXL SHM Arena}
\label{sec:cxl_shm_arena}
\textbf{Management of data objects.} For effective management of shared memory objects, we introduce the CXL SHM Arena, a CXL-based shared-memory management library integrated into MPI. While MPI traditionally uses POSIX shared memory (POSIX SHM) for intra-node IPC, the POSIX SHM API is unsuitable for managing CXL shared memory. 
POSIX SHM operates via memory-mapped files in a tmpfs filesystem (typically mounted at \texttt{/dev/shm}), where applications create POSIX SHM objects by naming files.
Typical usage involves: (1) using \texttt{shm\_open} to create/open a shared memory file in \texttt{/dev/shm}; (2) mapping this file into the caller's address space using \texttt{mmap}; (3) transferring data with \texttt{memcpy}; and (4) deleting data object with \texttt{shm\_unlink} once processes finish their operations.

\begin{figure}[tb!]
	\centering
         \includegraphics[width=1.0\linewidth]{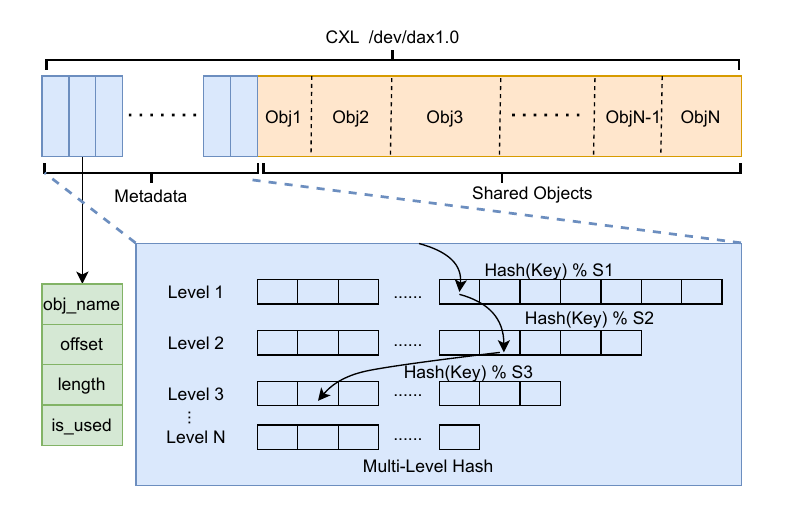}
	\caption{CXL SHM Arena.}
	\centering
	\label{fig:cxl_shm_arena} 
\end{figure}

However, POSIX SHM cannot manage CXL shared memory because of the following two reasons. First, unlike POSIX SHM, which uses a tmpfs filesystem, CXL shared memory is provided as a device in Device Direct Access (\texttt{devdax}) mode, also known as a \texttt{dax} device. Thus, we cannot create arbitrary SHM objects by creating uniquely-named files in a directory like \texttt{/dev/shm}. Although \texttt{mmap} can map different regions of a file by specifying file offset and size, the application must explicitly manage offsets. Also, the \texttt{dax} device requires the boundary-alignment mapping (e.g., 2MB alignment),  limiting the total number of SHM objects. 
Second, Linux cannot manage the lifecycle (creation, opening, closing, and deletion) of CXL SHM objects as it does with POSIX SHM objects. 

To address the above  limitation, we design the CXL SHM Arena. Figure~\ref{fig:cxl_shm_arena} generally depicts our design. The CXL SHM Arena maps the entire \texttt{dax} device into the process’s virtual address space and divides it into two regions: \texttt{metadata} and \texttt{shm\_objects}. The \texttt{metadata} region is organized as a fixed-size array of slots, each containing information about a shared memory object (e.g., usage status, object name, offset within the \texttt{dax} file, and object size). The \texttt{shm\_objects} region stores the actual shared data, allocated contiguously. 

\textbf{Multi-level hashing.} For efficient SHM object indexing, we employ a hash-based structure rather than a plain array to construct the \texttt{meta-data} region. We design the hash structure based on two requirements. First, it should effectively resolve hash collisions while maintaining a fixed capacity to avoid dynamic resizing. Second, it must support parallel insertions and searches. To meet these requirements, we use an existing fixed-capacity multi-level hashing scheme \cite{broder1990multilevel}, which is a representative of many multi-level hashing techniques~\cite{broder1990multilevel, broder2001using, zuo2019level, xiong2023dalea, zuo2018write, chen2020lock}.



Figure ~\ref{fig:cxl_shm_arena} illustrates the design, where buckets are organized into multiple levels, each with a distinct prime bucket-count to evenly distribute entries and reduce collisions. These levels are flattened into a contiguous array within the metadata region. The key lookup involves computing a hash value and examining each level in turn, continuing until the key is found or all levels are checked. The lookup can be parallelized across the levels, enhancing efficiency.


CXL SHM Arena creates SHM objects identified by a unique object name, which serves as the hash key. If the object name is not found in the metadata hashing, a new object is allocated in the free region of \texttt{shm\_objects}, with corresponding metadata recorded (including object name, address offset relative to the Arena base address, and size). The application subsequently opens an existing SHM object by searching its name in the metadata hashing to retrieve the offset and size, and then calculates the object's virtual address by adding the offset to the base address of the Arena.


\textbf{Famfs vs. CXL SHM Arena.} Famfs~\cite{famfs2024} is a shared-memory file system framework recently created by MICRON. Famfs is different from our design from three perspectives. \textit{First}, Famfs lives in the Linux kernel. In contrast, CXL SHM Arena is a user-space library without any kernel modification. \textit{Second}, Famfs uses a client/master architecture, and restricts the creation of shared files exclusively to the master node, limiting its applicability to MPI where any node may need to create SHM objects. In contrast, CXL SHM Arena supports the creation of SHM objects by arbitrary nodes. \textit{Third}, the SHM management APIs in CXL SHM Arena resemble those in the traditional POSIX SHM,  which facilitates easy integration into the application, whereas Famfs uses quite different APIs. 

\subsection{One-Sided MPI Communication over CXL}
MPI-3 introduces \texttt{MPI\_Win\_allocate\_shared} to create POSIX SHM-based RMA windows for one-sided communication. This call allocates a shared-memory segment for each MPI process on the same host. Those memory segments are laid out contiguously across all ranks. This means that the first address in the memory segment of the rank \(i\) is consecutive with the last address in the memory segment of the rank \(i-1\). This layout allows an MPI process to calculate the address of a RMA window using only local information.

\textbf{CXL SHM-based RMA window.} We extend \texttt{MPI\_Win\_allocate} \texttt{\_shared} to create CXL SHM-based RMA window for ranks across nodes. 
During MPI initialization, each rank maps the CXL \texttt{dax} device into its own virtual address space using \texttt{mmap}. This gives the rank a base address, called \texttt{meta\_addr}.  To create a CXL SHM-based RMA window for a specific communicator,  the root rank of the communicator creates a CXL SHM object in CXL SHM Arena.  The object size is equal to the number of ranks multiplied by the size of the RMA window for each rank.  The root rank then broadcasts the object name to all other ranks. Each rank uses this name to open the object (i.e., searching for the object in the CXL SHM Arena) and read the object metadata. Each rank gets the virtual address of its memory region by adding the metadata’s offset to \texttt{meta\_addr}. Since the memory segments are contiguous across ranks, each process can determine the virtual address of any other rank’s window using its own SHM object address and the other rank’s ID. 

To facilitate one-sided communication over CXL, we bypass network communication by allowing a rank to treat the communicating peer as a one on the same host. This allows RMA operations to proceed fully through CXL SHM. For \texttt{MPI\_Put()}, the origin rank directly writes data to the target rank’s RMA window from the origin rank's address space. For \texttt{MPI\_Get()}, the origin rank reads data directly from the target rank’s RMA window. 

\subsection{Two-Sided MPI Communication over CXL}
\label{sec:design:twosided}


In the current MPI implementations (e.g., MPICH and OpenMPI), the shared memory is commonly utilized for point-to-point (pt2pt) communication between processes on the same host. Within the shared memory, lock-free message queues are created as a channel for message passing. 
Each process maintains a receive queue, into which other processes on the same host can enqueue messages. Those queues are implemented as linked-list structures, containing head and tail pointers pointing to individual message cells. Those message cells store the actual message data and are linked together to form a \textit{message pool}. To send a message, the sending process copies the message header and data into a message cell and then enqueues the cell into the target process’s receive queue. To receive a message, the receiving process polls its own receive queue and retrieves incoming messages by dequeuing those message cells.

\textbf{CXL SHM-based message queue.} We extend the SHM-based pt2pt communication by enabling processes across different hosts (not just processes on the same host) to send messages via CXL SHM. Specifically, instead of creating a POSIX SHM-based message queue only for processes on the same host, we create a CXL SHM-based message queue accessible to all processes across the hosts. Similar to the SHM-based one-sided window creation, these queues are laid out contiguously within the CXL SHM, forming a message queue region. This contiguous arrangement allows any process to locate other processes' message queues based on the base address of the message queue region and the index of each message queue.


The current MPI implementations often create only one receive queue per host, employing either a Multi-Producers Single-Consumer (MPSC) or a Multi-Producers Multi-Consumer (MPMC) model. When using the sender-side message pool, an MPSC receive queue is employed, wherein multiple sender processes simultaneously dequeue available cells from their individual message pools and enqueue messages into the receiver’s queue (a single queue). In this scenario, multiple processes concurrently enqueue messages, but only a single receiver process dequeues messages from the receiver's queue. When using the receiver-side message pool, an MPMC queue is employed, wherein multiple sender processes simultaneously dequeue free message-cells from the receiver's message pool, fill the message cells with the message, and concurrently enqueue filled cells into the receiver's receive queue. In this scenario, multiple processes can simultaneously perform both enqueue and dequeue operations on the same queue.


Both the MPSC queue and MPMC queue are prone to race conditions when multiple processes attempt to enqueue or dequeue simultaneously. Although some MPI implementations (e.g., MPICH) employ lock-free queues using atomic operations to address the race condition, atomic operations are not effective in CXL SHM across multiple hosts (depicted in Section \ref{sec:design:consistency}).

To implement lock-free queues without relying on atomic operations, we design a Single-Producer Single-Consumer (SPSC) message ring queue. Specifically, we create a matrix of message ring queues for pt2pt communication. Unlike the traditional design where each process maintains a single receive queue for messages from all other processes, we establish separate ring queues for each pair of sender and receiver processes. These message queues follow the SPSC model and eliminate the need for atomic operations, as enqueue and dequeue operations are handled by a single producer and a single consumer, respectively. The collection of those message queues forms a message queue matrix, indexed by receiver rank ID and sender rank ID. To send a message, the sender locates the appropriate message queue and enqueues the message. To receive messages, the receiver polls ring queues associated with its ID and dequeues messages from each queue.

\subsection{Synchronization}
We optimize the most common synchronization mechanisms for CXL SHM, but keep other synchronization mechanisms unchanged. 

\textbf{Synchronization in the one-sided communication} is used to permit processes to access a target window. There are two common synchronization methods for the one-sided communication: 
Post-Start-Complete-Wait (PSCW) and Lock-Unlock. We leverage CXL SHM to reduce the one-sided synchronization overhead. PSCW enables a selected group of processes to participate in synchronization, using an access epoch and an exposure epoch. The target process invokes \texttt{Post} and \texttt{Wait} to define an exposure epoch, where it gives the origin process the availability of the RMA window. The origin processes invoke \texttt{Start} and \texttt{Complete} to define an access epoch, where RMA operations can be executed. The MPI implementation typically sends messages over the network to notify the origin and target processes of the epoch-status update. To reduce the PSCW overhead, we allocate a shared synchronization array during RMA window creation in CXL. Similar to the RMA window, the synchronization array is allocated contiguously across all processes. Each element in the array contains each process's  shared Post-Wait flag and shared Start-Complete flag. The origin process's Post-Wait flag is directly set by the target process and is reset by the origin. Conversely, the target process's Start-Complete flag is directly set by the origin process and is reset by the target. This method eliminates the synchronization message transfer over network.  

Similarly, Lock-Unlock synchronization can be optimized by placing the window lock in CXL SHM, thereby eliminating the network round trip for lock/unlock request-acknowledge. 

{\textbf{Synchronization in the two-sided communication} includes \texttt{MPI\_Wait} and \texttt{MPI\_Test}. The \texttt{MPI\_Wait} call blocks until a given request completes, whereas \texttt{MPI\_Test} returns immediately and indicates whether the request has finished. Both calls monitor the status of outstanding MPI send/receive requests and advance local progress for any pending message passing. Since the message paths have been optimized, we do not use CXL SHM for both calls due to little performance benefit.

\textbf{Initialization barrier} is used for the processes synchronization in the same node during the MPI initialization phase (e.g., socket creation,  broadcast table creation, and shared message queue creation). The current MPI library implements the initialization barrier using a sense-reversing algorithm where each MPI process maintains a local \texttt{sense} variable. As the MPI processes enter the barrier, they increment a shared counter; the last arriving process resets this counter and toggles a shared \texttt{wait} flag to indicate the barrier completion. The remaining processes spin-wait until the \texttt{wait} flag matches their local \texttt{sense}, after which all processes flip their \texttt{sense} in preparation for the next barrier. \name refactors the initialization barrier to support synchronization across nodes for the creation of shared message queue while avoiding atomic operations (i.e., atomic increment). Specifically, each process maintains a local barrier sequence number, and all processes share a barrier array to store those numbers. When entering the barrier, a process increments its local sequence number and updates the barrier array entry corresponding to its rank. The process then checks the sequence numbers of other ranks in the barrier  array and spin-waits until all other processes’ sequence numbers are equal to or greater than its own local sequence number.

\subsection{Memory Coherence}
\label{sec:design:consistency}
CXL SHM allows multiple hosts to access the same memory region. However, our CXL SHM lacks the cache coherence support, which causes the updates made by one host to be invisible to others. To maintain data coherence, we must rely on either non-cacheable memory sharing or a software-based cache coherence mechanism. While CXL 3.0 introduces support for Back-Invalidate (BI) to enforce hardware coherence across hosts, the anticipated growth in CXL memory size makes impractical a precise snoop filter that tracks each cache line \cite{jain2024memory}. In contrast, the software-based cache coherence has been shown to succeed to optimize performance in large shared address space~\cite{CXLReady_Micro23, ma2024hydrarpc}. We discuss non-cacheable memory sharing and software-based cache coherence as follows. 

\textbf{Non-cacheable memory sharing.} Modern x86 processors feature Memory Type Range Registers (MTRRs), which allow the CPU to access specific memory address ranges with different cacheability attributes, such as write-back, write-combining, write-through, write-protect, and uncachable. We configure \texttt{/proc/mtrr} to set the physical memory region of the CXL \texttt{dax} device as uncachable, thereby enabling CXL SHM accesses to bypass CPU caching. However, as demonstrated in Section \ref{sec:eval:cache}, uncachable access to CXL SHM results in significantly higher latency (exceeding 4,000 $\mu s$) when the data size exceeds 2 KB.

\textbf{Software-based cache coherence.} We use cache flush and memory fence. Cache flush instructions, such as \texttt{clwb}, \texttt{clflush}, and \texttt{clflushopt}, invalidate the cache line for specific addresses from  each level of the cache hierarchy in the coherence domain; if that cache line contains modified data, that data is written back to memory \cite{guide2011intel}. Memory fence instructions, such as store fence (\texttt{sfence}) and load fence (\texttt{lfence}), enforce ordering constraints on memory operations, ensuring that all operations issued prior to the fence are completed before those issued after the fence. We perform cache flush and memory fence when accessing the CXL SHM Arena, CXL SHM-based RMA windows, and CXL SHM-based message queues. Specifically, after every write, we perform a cache flush followed by a memory fence to guarantee that data is written back to memory. Conversely, before every read, we perform a memory fence followed by a cache flush to ensure that stale data, potentially held in caches or due to prefetching, is invalidated.

Given that uncachable access to CXL SHM leads to drastically higher latency for the large data size, 
we choose the software-based cache coherence. Additionally, we utilize non-temporal stores and loads when operating on integer data, such as the synchronization flags and the head/tail pointers of the message queue, to ensure that data is accessed directly from CXL SHM.

\subsection{Discussion}
\label{sec:design:limitation}


\name currently focuses on one-sided and two-sided communications across nodes. There are collective communications such as \texttt{Allgather}, \texttt{Allreduce}, and \texttt{ReduceScatter} that can  benefit from \textcolor{check}{CXL SHM}. Within an MPI library, these collective communications are often implemented using point-to-point communications based on certain algorithms (e.g., recursive doubling~\cite{arap2015adaptive} and Bruck's algorithm~\cite{fan2022optimizing}), hence the collective communications can directly benefit from \name. Nevertheless, there are multiple challenges of using \name for the collective communicators, such as communicator handling and integrating intra/inter-node point-to-point communications. We leave collective communications as future work.


\name exhibits performance benefits in inter-node communication, mostly for small message sizes (\textcolor{check}{16 KB or smaller}, shown in Section~\ref{sec:microbench}). When communicating using \textcolor{check}{CXL SHM}, the CPU must handle message transfers primarily using the \texttt{mov} instruction (or other load/store-like instructions) to copy data between the local buffer in main memory and the message queue or RMA window in \textcolor{check}{CXL SHM}. For small messages, the MPI communication bandwidth using  \textcolor{check}{CXL SHM} with 8 processes even outperforms that of TCP over Mellanox (CX-6 Dx) with 32 processes by up to 1.3$\times$, due to the lower latency of \textcolor{check}{CXL SHM}. However, for large messages, excessive CPU-based memory accesses lead to contention in the memory hierarchy, resulting in degraded communication performance. In contrast, the network-based approaches using high-end SmartNICs offer better scalability for large message transfers, as the actual data transmission occurs over the network with less CPU involvement. 


Using \textcolor{check}{CXL SHM} to accommodate RMA window for one-sided communication or the message buffer for two-sided communication can lead to performance loss, compared to using the node-local memory. This is because of the performance difference between the local memory and \textcolor{check}{CXL SHM} (e.g., 100 $ns$ vs. 2.2 $\mu s$ shown in Table~\ref{tab:memory_type}). However, such a performance difference is outweighed by the performance benefit of using the \textcolor{check}{CXL SHM} when the message size is small. The performance difference is more pronounced when the message size is large.


\subsection{More Implementation Details}
\label{sec:impl}

\begin{table}[!t]
    \centering
    \caption{CXL SHM Arena APIs for \name}
    \resizebox{\columnwidth}{!}{
    \begin{tabular}{|l|l|l|}
    
                \hline

        \textbf{API Function} & \textbf{Arguments} & \textbf{Description} \\

                \hline
        \addlinespace[4pt] 
        
        \hline

        \texttt{cxl\_shm\_init}       & --                          & Initialize and mmap the CXL SHM Arena \\
                \hline

        \texttt{cxl\_shm\_finalize}   & --                          & Clean up SHM arena and device memory \\
                \hline

        \texttt{cxl\_shm\_create}     & \texttt{name}, \texttt{size}, \texttt{*obj\_handle} & Create a new object with the specified size \\
                \hline

        \texttt{cxl\_shm\_open}       & \texttt{name}, \texttt{*obj\_handle}               & Open an existing object by name \\
                \hline

        \texttt{cxl\_shm\_destroy}    & \texttt{*obj\_handle}                             & Delete an object from the CXL SHM Arena \\
                \hline

        \texttt{cxl\_shm\_close}      & \texttt{*obj\_handle}                             & Close and release an object handle \\
                \hline

    \end{tabular}
    }
    \label{tab:cxl_shm_arena_api}
\end{table}

We implement \name by extending MPICH-4.2.3 ~\cite{mpich2}. In CXL SHM Arena, to enable \name to create millions of distinct CXL SHM objects, we configure a 10-level multi-level hash table and set the maximum number of slots at the level one to 200,000. Since the number of slots is constrained to be a prime number, the actual slot counts across the levels 1-10 range from 199,999 to 199,873. This results in a total of 1,999,260 slots across all levels.

To integrate CXL SHM Arena seamlessly with MPI, we design the APIs of CXL SHM Arena to resemble the POSIX SHM API, as shown in Table~\ref{tab:cxl_shm_arena_api}. This design ensures that generating SHM objects only requires API-level changes, without significantly modifying the logic flow of creating and broadcasting SHM objects in MPI.

To support memory consistency, we align each CXL SHM object to the cacheline size. This alignment facilitates efficient cache flushing and supports non-temporal memory accesses.

\section{Evaluation}
\label{sec:eval}

\begin{figure*}[tb!]
	\centering
         \includegraphics[width=1\linewidth]{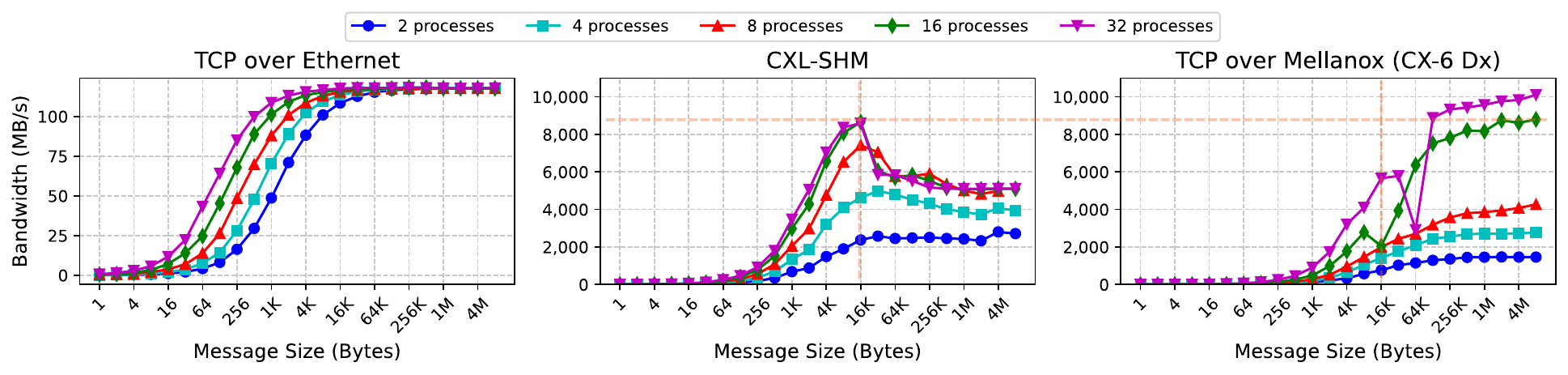}
	\caption{Bandwidth of one-sided MPI communication.}
	\centering
	\label{fig:osu_get_mbw} 
\end{figure*}

\begin{figure*}[tb!]
	\centering
         \includegraphics[width=1\linewidth]{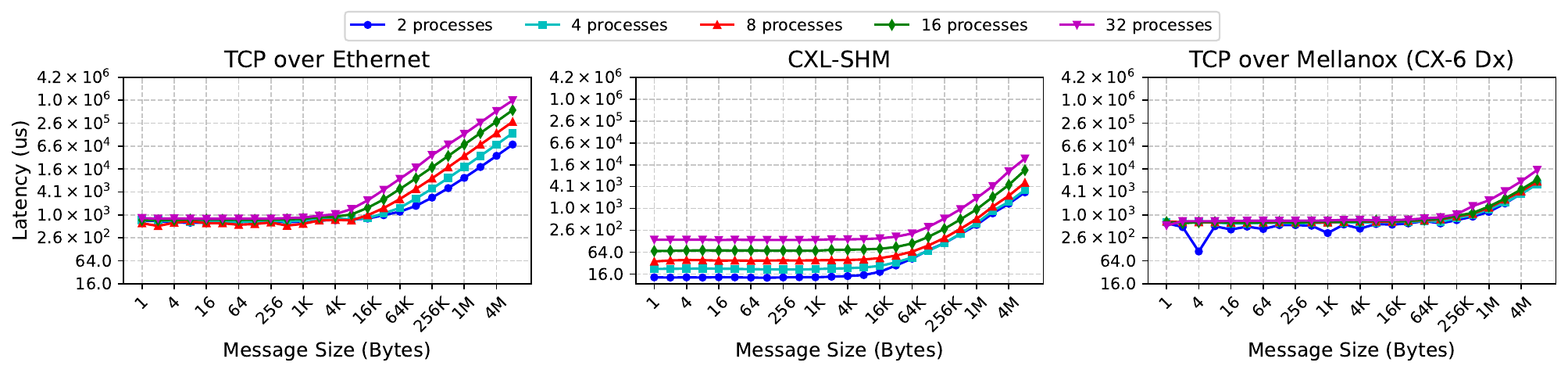}
	\caption{Latency of one-sided MPI communication.}
	\centering
	\label{fig:osu_get_multi_lat} 
\end{figure*}

\begin{figure*}[tb!]
	\centering
         \includegraphics[width=1\linewidth]{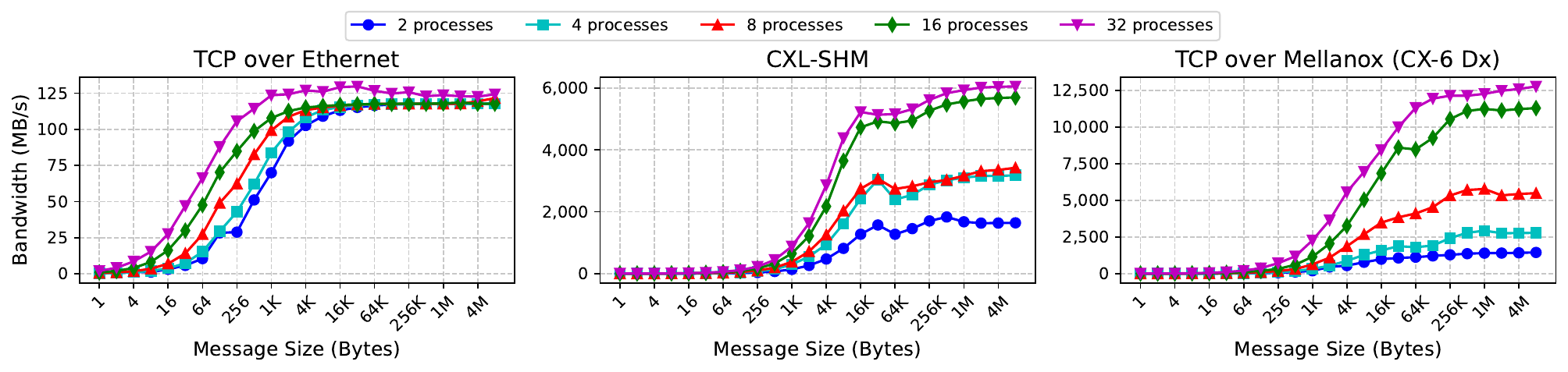}
	\caption{Bandwidth of two-sided MPI communication.}
	\centering
	\label{fig:osu_pt2pt_mbw} 
\end{figure*}

\begin{figure*}[tb!]
	\centering
         \includegraphics[width=1\linewidth]{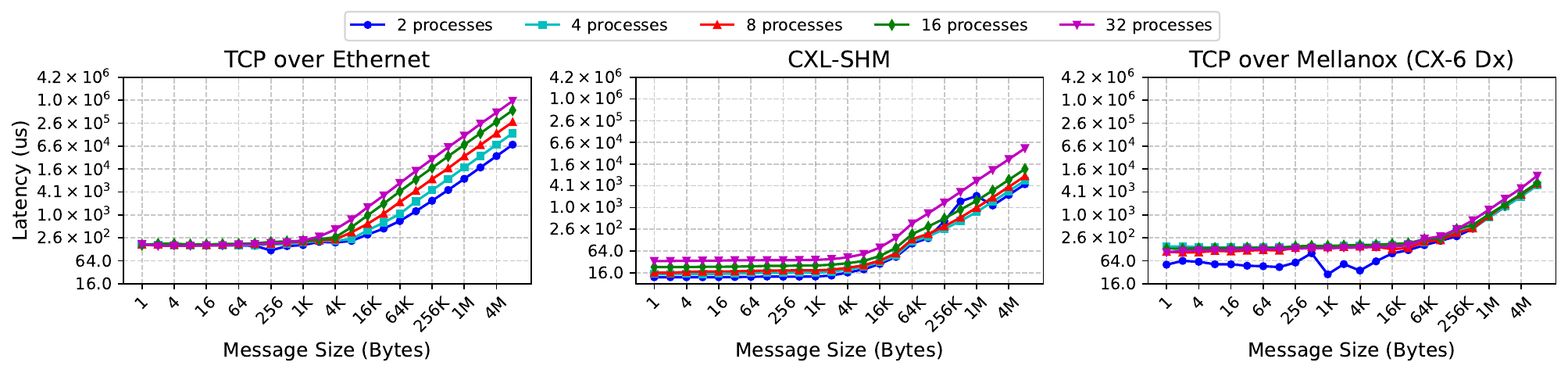}
	\caption{Latency of two-sided MPI communication.}
	\centering
	\label{fig:osu_pt2pt_multi_lat} 
\end{figure*}

\subsection{Evaluation Methodology}
\label{sec:eval:method}

\textbf{Evaluation platform.} We evaluate \name on two dual-socket servers. 
Each server is equipped with two Intel Xeon Gold 6530 processors (@2.10 GHz, 32 cores per socket). Each socket has 8×32 GB DDR5-5600 DIMM. 
We use MPICH-4.2.3 (CH4:OFI) and \name on Linux kernel 6.6.0-rc6, which is integrated with the CXL driver and DCD patches. 
To study the impact of communication across nodes and avoid the NUMA effect within each node, we disable NUMA balancing by setting \texttt{/proc/sys/kernel}\texttt{/numa\_balancing} to 0. 


\textbf{Real CXL Platform.} We use Niagara 2.0 \cite{Niagara20}, an FPGA-based CXL pooled memory platform. This platform can connect up to 4 nodes without a CXL switch. Each server connects to this platform via an MCIO x8 cable using PCIe 4.0 (16 GT/s per lane). The platform uses four DDR4-2400 DIMM channels, with a total capacity of 1TB and a bandwidth of 19.2 GB/s per channel. 


\textbf{Interconnect.} Each node uses a ConnectX-6 Dx SmartNIC (Mellanox Technologies MT2892 Family), supporting Ethernet and RoCEv2.  Two nodes are also connected via another standard Ethernet NIC with limited performance (see ``TCP over Ethernet'' in Table~\ref{tab:memory_type}).


\textbf{Baselines.} We compare our approach (CXL SHM-based communication) with TCP over Ethernet and TCP over Mellanox (CX-6 Dx). We do not compare with RoCEv2 over Mellanox (CX-6 Dx and CX3) and InfiniBand (Melanox CX-6), as \textcolor{check}{CXL SHM} is not competitive with them in terms of latency and bandwidth (see Table~\ref{tab:memory_type}). 

\textbf{Simulator for scalability study.} We use SimGrid~\cite{CASANOVA2025103125}, an event-based simulator that can study MPI applications with user-specified interconnect latency and bandwidth. 
\textcolor{check}{Because our CXL platform is constrained to connecting only four hosts, simulation is necessary to study scalability beyond this hardware limitation.}


\subsection{Microbenchmarks}
\label{sec:microbench}
We use the OSU Micro-Benchmark suite (OMB)~\cite{osu_omb} to compare performance (i.e., latency and bandwidth) across two nodes using CXL SHM, TCP over Ethernet, and TCP over Mellanox (CX-6 Dx). To evaluate the performance of one-sided communication under multiple processes, we extend the OMB's benchmark for one-sided communication, which originally only supports two processes, to support any number of MPI origin and target processes. To evaluate the performance of two-sided communication, we change the size per message cell to 64KB to improve performance (see Section~\ref{sec:eval:cell_size}). We evaluate the message size from 1B to 8MB, representing small, medium, and large message sizes~\cite{osu_omb}. 
For one-sided communication, given $n$ MPI processes, half of them is used as the origin processes, and the other half is used as the target processes.


\textbf{Bandwidth of one-sided communication.} See Figure~\ref{fig:osu_get_mbw}. 

\underline{TCP over Ethernet vs. CXL SHM. } 
CXL SHM outperforms TCP over Ethernet by up to 71.6$\times$. With CXL SHM, the bandwidth saturates at around 8,600 MB/s with 16 processes when the message size reaches 16 KB. With TCP over Ethernet, adding more processes can saturate the interconnect bandwidth using smaller message sizes, but the bandwidth curves converge near 120 MB/s. 

\underline{TCP over Mellanox (CX-6 Dx) vs. CXL SHM.} For message sizes of 16 KB or smaller, CXL SHM outperforms TCP over Mellanox (CX-6 Dx) by up to 3.7$\times$. CXL SHM achieves higher bandwidth even using fewer processes compared to TCP over Mellanox (CX-6 Dx). For example, CXL SHM achieves 7,420 MB/s at 16 KB using 8 processes, whereas TCP over Mellanox only reaches 5,660 MB/s at 16 KB even with 32 processes. The high bandwidth of CXL SHM for small messages stems from the low latency of CPU‐based accesses to small data. Beyond 16 KB, however, the CXL SHM bandwidth declines because concurrent accesses to large data from multiple processes induce contention on the memory hierarchy.

For messages larger than 16 KB, TCP over Mellanox (CX-6 Dx) surpasses CXL SHM when using 16 or 32 processes. Using 32 processes, TCP over Mellanox (CX-6 Dx) allows the bandwidth to climb to 10,150 MB/s as the message size grows, which is near the theoretical peak bandwidth (100 Gb/s). Although TCP over Mellanox (CX-6 Dx) initially requires the CPU to package the message, the subsequent network transfer proceeds without further CPU involvement. In contrast, CXL SHM relies on CPU \texttt{mov} instruction for the entire one‐sided message passing, continuously involving the CPU. Consequently, TCP over Mellanox (CX-6 Dx) has the advantage for larger data transfers, because of less involvement of the CPU. 


\textbf{Latency of one-sided communication}. See Figure \ref{fig:osu_get_multi_lat}.

\underline{TCP over Ethernet vs. CXL SHM.}
CXL SHM outperform TCP over Ethernet by up to 49.4$\times$. With CXL SHM, the latency remains consistently low (around 12 $\mu s$) for message sizes from 1 B to 16 KB, and increases  modestly as more processes are used. Conversely, TCP over Ethernet experiences latency in the hundreds of microseconds (approximately 630 $\mu s$) even for small messages.  Increasing the process count from 2 to 32 shifts the curve only slightly because more processes make the link busier. Ethernet’s inherent limitations (e.g., multiple rounds of data copy and deep software stack) cap overall performance.

\underline{TCP over Mellanox (CX-6 Dx) vs. CXL SHM. }
For message sizes of 256 KB or smaller, CXL SHM achieves up to 48.3$\times$ lower latency than TCP over Mellanox (CX-6 Dx). Specifically, the latency for TCP over Mellanox (CX-6 Dx) hovers around 620 $\mu s$, while CXL SHM has a minimal latency of approximately 12 $\mu s$. 

For messages larger than 256 KB, TCP over Mellanox (CX-6 Dx) outperforms CXL SHM because its latency increases more slowly for larger data. This advantage stems from offloaded network transfers that do not require continuous CPU intervention. In contrast, the latency with CXL SHM increases proportionally as the data size grows when the data size is larger than 16 KB. This growth is mainly due to contention from concurrent accesses to large data from multiple processes. This observation is aligned with the observations from the one‐sided bandwidth evaluation.

\textbf{Bandwidth of two-sided communication.} See Figure \ref{fig:osu_pt2pt_mbw}.

\underline{TCP over Ethernet vs. CXL SHM. }
CXL SHM outperforms TCP over Ethernet by up to 48.2$\times$. With CXL‐SHM, the bandwidth saturates at around 6,050 MB/s. With TCP over Ethernet, the bandwidth exhibits a trend similar to its one‐sided counterpart and saturates at about 120 MB/s due to the throughput ceiling of Ethernet interface.   

\underline{TCP over Mellanox (CX-6 Dx)  vs. CXL SHM. }
TCP over Mellanox (CX-6 Dx) outperforms CXL SHM by up to 2.1$\times$ when using more than 4 processes. With CXL SHM, the peak bandwidth for the two-sided communication (about 6,050 MB/s) is approximately 30\% lower than its one‐sided counterpart (8,600 MB/s). This occurs because each two‐sided operation involves two message transfers: one from the sender’s local buffer to the message queue in CXL SHM, and another from that message queue to the receiver’s local buffer. Consequently, the memory access contention doubles at the DIMM and memory bus level in the CXL platform, which leads to lower bandwidth. 
With TCP over Mellanox, the scalability is stronger than with CXL SHM, as adding more processes continues to boost bandwidth. The bandwidth reaches beyond 12,500 MB/s at 32 processes for large message sizes.

\textbf{Latency of two-sided communication.} See Figure \ref{fig:osu_pt2pt_multi_lat}.

\underline{TCP over Ethernet vs. CXL SHM. }
CXL SHM outperforms TCP over Ethernet by up to 13.7$\times$. With TCP over Ethernet, the two-sided communication achieves lower latency (approximately 160 $\mu s$) than its one-sided counterpart (630 $\mu s$) due to extra synchronization round‐trips in the one‐sided communication. Nevertheless, the overall trend is similar to that of the one‐sided communication, with latency constrained by Ethernet.  In contrast, CXL SHM maintains a latency of around 12 $\mu s$, highlighting its advantage for latency-sensitive communications.

\underline{TCP over Mellanox (CX-6 Dx)  vs. CXL SHM. }
For message sizes smaller than 64 KB, CXL SHM achieves up to 9.6$\times$ lower latency, compared to TCP over Mellanox (CX-6 Dx).  Once the message size reaches 64 KB or larger, the latency with CXL SHM increases in proportion to the message size. This is due to the limited cell size (64 KB) in the message queue, which cannot accommodate a 64 KB message plus its header. As a result, a larger message is split into smaller chunks and sent sequentially through the message queue, causing a linear increase in latency.

With TCP over Mellanox (CX-6 Dx), for messages smaller than 256 KB, the two‐sided communication also exhibits lower latency (around 55 $\mu s$), compared to its one‐sided counterpart (around 620 $\mu s$). Beyond 256 KB, however, the latency scales linearly with the message size, reflecting bandwidth saturation and aligning with the results for two‐sided communication.

\subsection{Message Size Threshold for CXL SHM-based Two-Sided Communication}
\label{sec:eval:cell_size}
As described in Section \ref{sec:design:twosided}, the two‐sided communication uses a shared message queue in which head and tail pointers point to “message cells” that store both message headers and message payloads. Specifically, a sender copies data from its send buffer into one of those cells, while a receiver copies data out of the cell and into its receive buffer. MPI classifies messages whose size is no larger than a cell’s capacity as short messages; a longer message is split into multiple cell-sized chunks sent sequentially. Consequently, the size of message cell impacts parallelism in data transfers.


Figure \ref{fig:osu_pt2pt_cell_size} shows that with the MPI default cell size (16 KB),  the highest bandwidth achieved is 3600 MB/s (when the message size is 8 KB and the number of processes is 32). Beyond 16 KB, the message must be split, causing a drop in bandwidth. Increasing the cell size to 32 KB raises bandwidth to 3900 MB/s (when the message size is 16 KB and the number of processes is 32). 


Setting the message cell size to 64 KB improves peak bandwidth to approximately 6000 MB/s for messages larger than 64 KB with 32 processes. However, increasing the cell size beyond 64 KB does not continue to improve bandwidth, indicating an upper limit to the benefits of using a larger message cell.

\begin{figure}[tb!]
	\centering
         \includegraphics[width=0.9\linewidth]{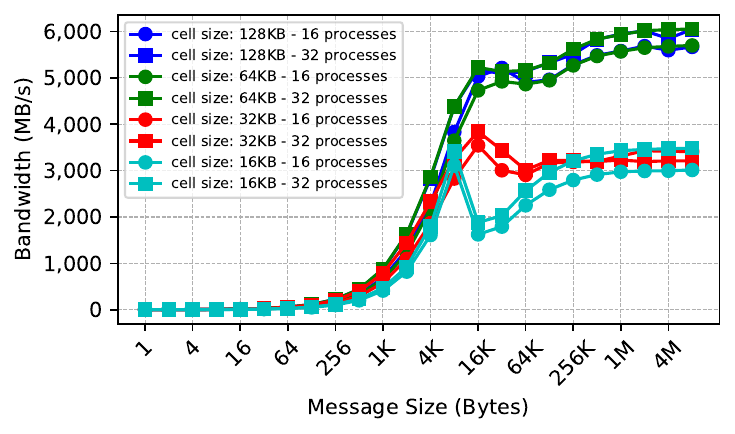}
	\caption{Bandwidth of two-sided communication with various sizes of message cells.}
	\centering
	\label{fig:osu_pt2pt_cell_size} 
\end{figure}

\subsection{Scalability Study}

\begin{figure}[tb!]
	\centering
         \includegraphics[width=1\linewidth]{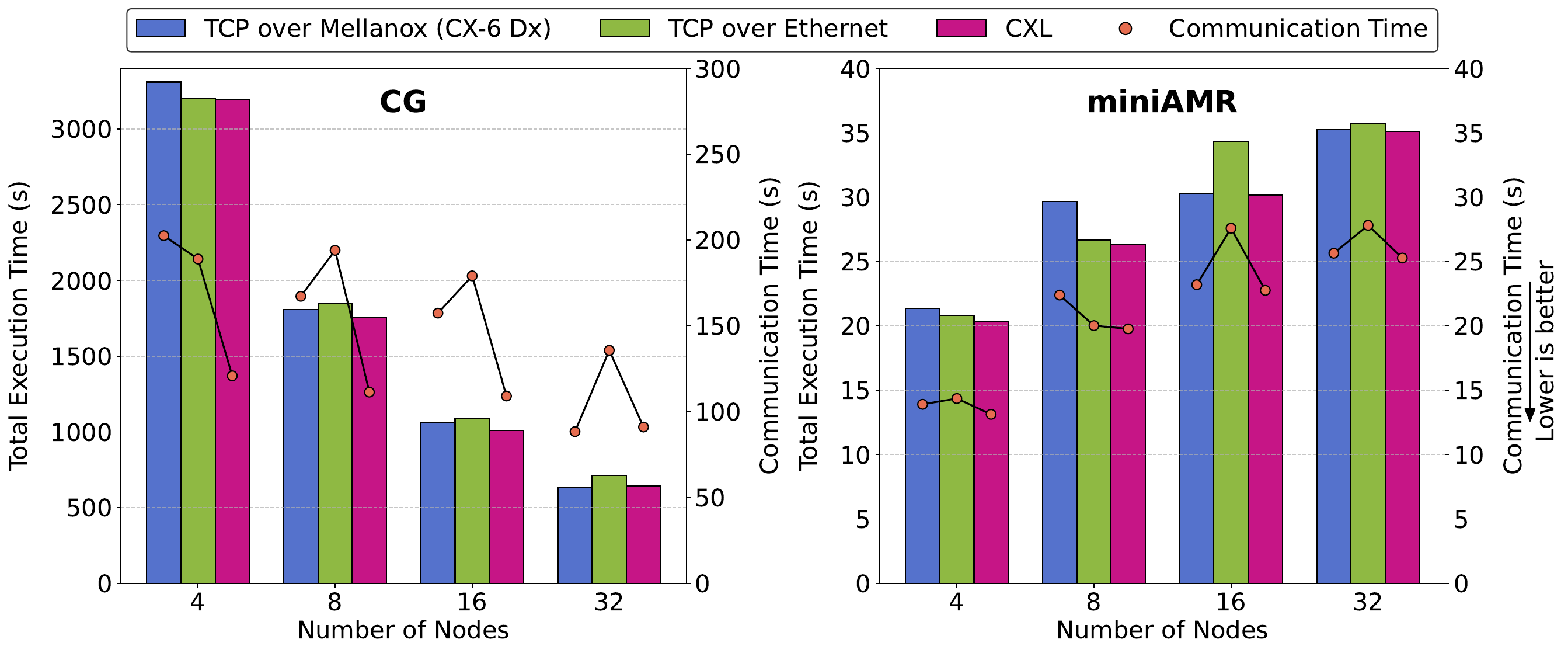}
	\caption{Strong scaling with CG and miniAMR.}
	\centering
	\label{fig:scalability_cg} 
\end{figure}

\begin{figure}[tb!]
	\centering
         \includegraphics[width=0.9\linewidth]{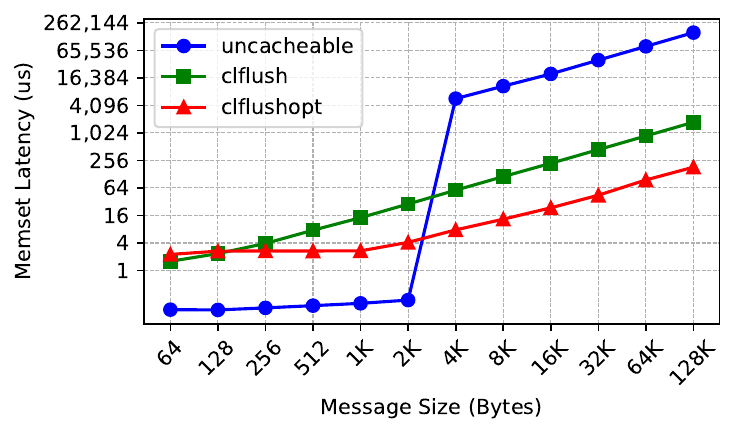}
	\caption{Memset latency with uncacheable memory and cacheable memory (plus cache flushing overhead).}
	\centering
	\label{fig:memset_latency}
\end{figure}

We use SimGrid simulator to perform strong scaling evaluation, because our CXL platform does not support large-scale evaluation. 
The latency and bandwidth of inter-node and intra-node are configured based on our evaluation results in Section~\ref{sec:microbench}. We use eight MPI processes per node during the evaluation.

We use CG (a conjugate gradient solver) from NAS Parallel Benchmark (NPB) Suite~\cite{doi:10.1177/109434209100500306} 3.4. For CG, we use Class D as input. We also use miniAMR~\cite{miniAMR} (a proxy application for  adaptive mesh refinement) with the input block size 4 in $x,y,z$ directions. We choose CG and miniAMR for evaluation, because their communications are dominated by MPI two-sided communications \cite{osti_10450791,CG_p2p}.

\textbf{CG results.} CXL SHM outperforms other two network-based approaches because of its shorter communication time. Specifically, the communication time using CXL SHM is shorter than that of TCP over Mellanox (CX-6 Dx) and TCP over Ethernet by 25.3\% and 37.6\%, respectively. This reduction is attributed to the lower inter-node latency provided by CXL SHM. As the number of nodes increases, the communication gap between CXL SHM and TCP over Mellanox (CX-6 Dx) narrows, since the higher bandwidth of Mellanox (CX-6 Dx) benefits message transfers at larger scales. However, because communication accounts for a small portion (less than 15\%) of total  execution time, the overall performance difference is small.

\textbf{miniAMR results.} CXL SHM achieves an average speedup of 4\% and 4.7\% in total execution time compared to TCP over Mellanox (CX-6 Dx) and TCP over Ethernet, respectively. Unlike CG, miniAMR has a higher communication portion  (more than 62\%) of total
execution time. As the number of nodes increases, the communication time grows while the computation time remains steady, since miniAMR assigns each process a fixed number of grid blocks and each process performs a constant computational workload. The reduced communication time with CXL SHM—4.9\% lower than Mellanox and 9.8\% lower than Ethernet—directly contributes to the performance gain. Additionally, TCP over Ethernet outperforms TCP over Mellanox (CX-6 Dx) at 8 nodes or fewer due to its slightly lower latency (16~$\mu$s vs. 18~$\mu$s), but performs worse beyond 8 nodes because of its significantly lower bandwidth (117.8 MB/s vs. 11.5 GB/s). At small scales, communication time is latency-dominated, while at larger scales, bandwidth becomes the limiting factor.


\subsection{Cache Coherence}
\label{sec:eval:cache}
To maintain CXL memory coherence, we can either use non-cacheable memory sharing or a software-based cache coherence mechanism, discussed in Section~\ref{sec:design:consistency}.
To compare the two approaches, we use our micro-benchmark (see Section~\ref{sec:bg}) to measure latency across data sizes ranging from 64 B to 128 KB.
For the software-based cache coherence, the \texttt{memset} latency includes caching flushing overhead.

For the uncacheable approach, we use MTRR to mark the physical memory region of the CXL \texttt{dax} device as uncacheable before running the benchmark, causing all subsequent memory accesses to bypass the CPU cache. We then execute \texttt{memset} without performing cache flushes. For the \texttt{clflush} approach, using MTRR the CXL SHM region is configured to be write-back cacheable. We execute \texttt{memset} followed by \texttt{clflush} and an \texttt{sfence} to ensure proper memory ordering. 
We use the same approach when evaluating \texttt{clflushopt}. 
Compared to \texttt{clflush}, \texttt{clflushopt} can flush multiple cache lines in parallel.

Figure \ref{fig:memset_latency} shows the results. We observe that when the data size is larger than 2 KB, the uncacheable accesses experience  256$\times$ higher latency, compared to using \texttt{clflush} and \texttt{clflushopt}. Moreover, \texttt{clflushopt} outperforms \texttt{clflush} by up to 4$\times$ when the data size is larger than 64 B. In particular, when the data size ranges from 1 B to 64 B, both \texttt{clflush} and \texttt{clflushopt} exhibit similar latencies (around 2 $\mu s$ to 3 $\mu s$), since the data size does not exceed a single cache line, requiring only one cache flush. Beyond 64 B, multiple cache flushes are necessary, and \texttt{clflushopt} performs better because of its efficient parallel flushing.

When the data size is larger than 2 KB, uncacheable accesses lead to sudden latency spikes (exceeding 4096 $\mu s$). This behavior arises from the Maximum Payload Size (MPS) limitation in PCIe, which is typically between 128 B and 4096 B. Once the data size exceeds the MPS, the PCIe link splits it into multiple Transaction Layer Packets, and transfers them sequentially, increasing latency.



\section{Related Work}


\textcolor{check}{\textbf{Intra-node MPI communication over shared memory.}} 
\textcolor{check}{Optimizing intra-node MPI communication within shared memory has been extensively studied ~\cite{10.1145/2462902.2462903,10.1007/978-3-540-87475-1_21,4663761,10.1007/978-3-031-73370-3_6, 10898940, memoManaMPI_EuroMPI17, white2020optimizing, adaptiveMPI}. For example,}
Adam et al.~\cite{10.1007/978-3-031-73370-3_6} reveal significant bandwidth differences between cores and sockets in a node and demonstrate that current MPI implementations may not fully utilize the intra-node bandwidth
potential. 
Walia et al.~\cite{10898940} leverage OpenMP for accelerating MPI\_Allreduce in deep learning workloads. Cho et al. \cite{memoManaMPI_EuroMPI17} improve MPI bandwidth performance through on-package memory and huge page support. White and Kale~\cite{white2020optimizing} optimize point-to-point communication in AMPI~\cite{adaptiveMPI} by their locality-aware design in shared memory. \textcolor{check}{Unlike previous works focusing on intra-node communication, our study optimizes inter-node MPI communication over CXL SHM.}

\textbf{Evaluation of CXL.}
\textcolor{check}{Recent CXL memory evaluations provide insights into this emerging memory technology \cite{liu2024dissectingcxlmemoryperformance,CXLReady_Micro23,yang2025architecturalimplicationscxlenabledtiered,wang2025cxl_performance,CXLReady_Micro23,PerfCostCXL_Eurosys24,memsys_cxl_switch_2024}.}
\textcolor{check}{For example, Sun et al. \cite{CXLReady_Micro23} demonstrate important differences between real and emulated CXL memory, necessitating reconsideration of previous assumptions on the CXL hardware. Tang et al. \cite{PerfCostCXL_Eurosys24} evaluate CXL performance and design an Abstract Cost Model to estimate the cost-benefit of using CXL memory in datacenters. }
\textcolor{check}{Wang et al. \cite{wang2025cxl_performance} provide a comprehensive evaluation of how CXL interacts with GPU in LLM training and inference.}

\textbf{CXL shared memory management.}
Famfs~\cite{famfs2024,10898426} from Micron is an open source file system for scale-out disaggregated shared memory across multiple nodes. We discuss its difference from \name in Section~\ref{sec:cxl_shm_arena}. 
\textcolor{check}{Zhang et al. \cite{FailureCXL_sosp23} develop a failure-resilient CXL SHM system using reference counting to tolerate partial client failures.}

\textbf{CXL for MPI.}
OMB-CXL~\cite{OMBCXL2024} introduces a micro-benchmark suite for evaluating MPI communications over CXL SHM based on emulation. 
\textcolor{check}{Ahn et al. ~\cite{10825804} present the first MPI over CXL SHM implementation beyond benchmarking,  utilizing CXL SHM to accelerate MPI\_Allgather across nodes. However, this work addresses only a single collective operation and is based on emulation.}
\textcolor{check}{In contrast, our work is the first to optimize MPI point-to-point communication (both one-sided and two-sided) on real CXL Pooled Memory Platforms, with novel approaches to build DAX abstraction, MPI synchronization, and memory coherence. }

\textbf{Applications of CXL memory.} CXL has been explored by various application domains. For databases, Ahn et al. \cite{ahn2022enabling} leverage CXL memory expansion to enhance in-memory database systems. For AI workloads, Xie et al. \cite{SmartQuant} introduce SmartQuant, a CXL-based AI model store supporting runtime weight quantization to reduce inference latency. For approximate nearest neighbor search (ANNS), Jang et al. \cite{CXLANNSATC23} propose CXL-ANNS, which utilizes CXL-based memory disaggregation and parallel search techniques to efficiently handle large datasets. 

\textcolor{check}{\textbf{Tiered memory systems.} Apart from serving as a shared memory pool, CXL can also be used to build a tiered memory system for a host. With a tiered memory system, various memory components~\cite{wang2020characterizing, xiang2022characterizing, bergman2022reconsidering, dragojevic2014farm, gu2017efficient, abulila2019flatflash, van2019hoti, sharma2022compute,gpu_dp_micro14,wang2025cxl_performance} exhibit differences in latency, bandwidth, capacity, and cost. A range of solutions ~\cite{Maruf2022MULTICLOCKDT, Kumar2021RadiantEP, Hildebrand2020AutoTMAT, Li2022TransparentAL, Ren2021SentinelET, Wang2019PantheraHM, Dulloor2016DataTI, Raybuck2021HeMemST, Kannan2017HeteroOSO, Kim2021ExploringTD, Weiner2022TMOTM, sc18:wu, unimem:sc17, ics21:warpx, wahlgren2023quantitative, ppopp23:merchandiser,eurosys24:mtm,atc24_hm,micro18:pim,8048928,neurips20:hm-ann,ics21:athena,ppopp21:sparta,luo:NGS,hpca25:dlrm,hpca25:buffalo,betty:asplos23,hpdc16:wu,wang2025cxl_performance} have been proposed to leverage tiered memory systems to improve application performance.}
\section{Conclusions}
In this paper, we explore how to use CXL memory sharing for MPI point-to-point communication across nodes. Our study reveals the performance potential of using CXL memory sharing, compared to certain interconnects. We also study how to customize the MPI library to use CXL memory sharing. Our work opens a door for using CXL techniques in HPC.
\begin{acks}
\textcolor{check}{This work was partially supported by U.S. National Science Foundation (2104116, 2316202 and 2348350). We would like to thank the anonymous reviewers for their feedback on the paper.}
\end{acks}

\bibliographystyle{ACM-Reference-Format}
\bibliography{sherry,bin,li,jie}

@inproceedings{memoManaMPI_EuroMPI17,
author = {Cho, Joong-Yeon and Jin, Hyun-Wook and Nam, Dukyun},
title = {Enhanced memory management for scalable MPI intra-node communication on many-core processor},
year = {2017},
isbn = {9781450348492},
publisher = {Association for Computing Machinery},
address = {New York, NY, USA},
url = {https://doi.org/10.1145/3127024.3127035},
doi = {10.1145/3127024.3127035},
booktitle = {Proceedings of the 24th European MPI Users' Group Meeting},
articleno = {10},
numpages = {9},
location = {Chicago, Illinois},
series = {EuroMPI '17}
}

@inproceedings{mpich2,
author = {Gropp, William},
title = {MPICH2: A New Start for MPI Implementations},
year = {2002},
isbn = {3540442960},
publisher = {Springer-Verlag},
address = {Berlin, Heidelberg},
booktitle = {Proceedings of the 9th European PVM/MPI Users' Group Meeting on Recent Advances in Parallel Virtual Machine and Message Passing Interface},
pages = {7}
}

@misc{liu2024dissectingcxlmemoryperformance,
      title={Dissecting CXL Memory Performance at Scale: Analysis, Modeling, and Optimization}, 
      author={Jinshu Liu and Hamid Hadian and Hanchen Xu and Daniel S. Berger and Huaicheng Li},
      year={2024},
      eprint={2409.14317},
      archivePrefix={arXiv},
      primaryClass={cs.OS},
      url={https://arxiv.org/abs/2409.14317}, 
}

@misc{yang2025architecturalimplicationscxlenabledtiered,
      title={Architectural and System Implications of CXL-enabled Tiered Memory}, 
      author={Yujie Yang and Lingfeng Xiang and Peiran Du and Zhen Lin and Weishu Deng and Ren Wang and Andrey Kudryavtsev and Louis Ko and Hui Lu and Jia Rao},
      year={2025},
      eprint={2503.17864},
      archivePrefix={arXiv},
      primaryClass={cs.AR},
      url={https://arxiv.org/abs/2503.17864}, 
}

@inproceedings{CXLReady_Micro23,
author = {Sun, Yan and Yuan, Yifan and Yu, Zeduo and Kuper, Reese and Song, Chihun and Huang, Jinghan and Ji, Houxiang and Agarwal, Siddharth and Lou, Jiaqi and Jeong, Ipoom and Wang, Ren and Ahn, Jung Ho and Xu, Tianyin and Kim, Nam Sung},
title = {Demystifying CXL Memory with Genuine CXL-Ready Systems and Devices},
year = {2023},
isbn = {9798400703294},
publisher = {Association for Computing Machinery},
address = {New York, NY, USA},
url = {https://doi.org/10.1145/3613424.3614256},
doi = {10.1145/3613424.3614256},
booktitle = {Proceedings of the 56th Annual IEEE/ACM International Symposium on Microarchitecture},
pages = {105–121},
numpages = {17},
location = {Toronto, ON, Canada},
series = {MICRO '23}
}

@inproceedings{PerfCostCXL_Eurosys24,
author = {Tang, Yupeng and Zhou, Ping and Zhang, Wenhui and Hu, Henry and Yang, Qirui and Xiang, Hao and Liu, Tongping and Shan, Jiaxin and Huang, Ruoyun and Zhao, Cheng and Chen, Cheng and Zhang, Hui and Liu, Fei and Zhang, Shuai and Ding, Xiaoning and Chen, Jianjun},
title = {Exploring Performance and Cost Optimization with ASIC-Based CXL Memory},
year = {2024},
isbn = {9798400704376},
publisher = {Association for Computing Machinery},
address = {New York, NY, USA},
url = {https://doi.org/10.1145/3627703.3650061},
doi = {10.1145/3627703.3650061},
booktitle = {Proceedings of the Nineteenth European Conference on Computer Systems},
pages = {818–833},
numpages = {16},
location = {Athens, Greece},
series = {EuroSys '24}
}

@misc{famfs2024,
  author = {{CXL Micron Research Kit}},
  title = {{Famfs Shared Memory Filesystem Framework}},
  howpublished = {\url{https://github.com/cxl-micron-reskit/famfs}},
  year = {2024},
  note = {Accessed: 2025-04-05}
}

@inproceedings{FailureCXL_sosp23,
author = {Zhang, Mingxing and Ma, Teng and Hua, Jinqi and Liu, Zheng and Chen, Kang and Ding, Ning and Du, Fan and Jiang, Jinlei and Ma, Tao and Wu, Yongwei},
title = {Partial Failure Resilient Memory Management System for (CXL-based) Distributed Shared Memory},
year = {2023},
isbn = {9798400702297},
publisher = {Association for Computing Machinery},
address = {New York, NY, USA},
url = {https://doi.org/10.1145/3600006.3613135},
doi = {10.1145/3600006.3613135},
booktitle = {Proceedings of the 29th Symposium on Operating Systems Principles},
pages = {658–674},
numpages = {17},
location = {Koblenz, Germany},
series = {SOSP '23}
}

@inproceedings{OMBCXL2024,
author = {Tran, Tu and Abduljabbar, Mustafa and Ahn, Hooyoung and Kim, Seonyoung and Park, Yoomi and Han, Woojong and Ahn, Shinyoung and Subramoni, Hari and Panda, Dhabaleswar K.},
title = {OMB-CXL: A Micro-Benchmark Suite for Evaluating MPI Communication Utilizing Compute Express Link Memory Devices},
year = {2024},
isbn = {9798400704192},
publisher = {Association for Computing Machinery},
address = {New York, NY, USA},
url = {https://doi.org/10.1145/3626203.3670533},
doi = {10.1145/3626203.3670533},
booktitle = {Practice and Experience in Advanced Research Computing 2024: Human Powered Computing},
articleno = {27},
numpages = {8},
location = {Providence, RI, USA},
series = {PEARC '24}
}

@misc{SmartQuant,
      title={SmartQuant: CXL-based AI Model Store in Support of Runtime Configurable Weight Quantization}, 
      author={Rui Xie and Asad Ul Haq and Linsen Ma and Krystal Sun and Sanchari Sen and Swagath Venkataramani and Liu Liu and Tong Zhang},
      year={2024},
      eprint={2407.15866},
      archivePrefix={arXiv},
      primaryClass={cs.LG},
      url={https://arxiv.org/abs/2407.15866}, 
}

@inproceedings {CXLANNSATC23,
author = {Junhyeok Jang and Hanjin Choi and Hanyeoreum Bae and Seungjun Lee and Miryeong Kwon and Myoungsoo Jung},
title = {{CXL-ANNS}: {Software-Hardware} Collaborative Memory Disaggregation and Computation for {Billion-Scale} Approximate Nearest Neighbor Search},
booktitle = {2023 USENIX Annual Technical Conference (USENIX ATC 23)},
year = {2023},
isbn = {978-1-939133-35-9},
address = {Boston, MA},
pages = {585--600},
url = {https://www.usenix.org/conference/atc23/presentation/jang},
publisher = {USENIX Association},
month = jul
}

@INPROCEEDINGS{miniAMR,
  author={Vaughan, Courtenay T. and Barrett, Richard F.},
  booktitle={2015 IEEE International Conference on Cluster Computing}, 
  title={Enabling Tractable Exploration of the Performance of Adaptive Mesh Refinement}, 
  year={2015},
  volume={},
  number={},
  pages={746-752},
  doi={10.1109/CLUSTER.2015.129}}

@article{osti_10450791,
place = {Country unknown/Code not available}, title = {Understanding the use of message passing interface in exascale proxy applications}, url = {https://par.nsf.gov/biblio/10450791}, DOI = {10.1002/cpe.5901}, abstractNote = {Summary The Exascale Computing Project (ECP) focuses on the development of future exascale‐capable applications. Most ECP applications use the message passing interface (MPI) as their parallel programming model with mini‐apps serving as proxies. This paper explores the explicit usage of MPI in such ECP proxy applications. We empirically analyze 14 proxy applications from the ECP Proxy Apps Suite. We use the MPI profiling interface (PMPI) to collect MPI usage patterns in ECP proxy apps. Our analysis shows that a small subset of features from MPI is commonly used in the proxies of exascale‐capable applications, even when they reference third‐party libraries. This study is intended to provide a better understanding of the use of MPI in current exascale applications. The findings can help focus software investments made for exascale systems in the MPI middleware including optimization, fault‐tolerance, tuning, and hardware‐offload.}, journal = {Concurrency and Computation: Practice and Experience}, volume = {33}, number = {14}, publisher = {Wiley Blackwell (John Wiley & Sons)}, author = {Sultana, Nawrin and Rüfenacht, Martin and Skjellum, Anthony and Bangalore, Purushotham and Laguna, Ignacio and Mohror, Kathryn}, }

@inproceedings{CG_p2p,
author = {Small, Matthew and Yuan, Xin},
title = {Maximizing MPI point-to-point communication performance on RDMA-enabled clusters with customized protocols},
year = {2009},
isbn = {9781605584980},
publisher = {Association for Computing Machinery},
address = {New York, NY, USA},
url = {https://doi.org/10.1145/1542275.1542320},
doi = {10.1145/1542275.1542320},
booktitle = {Proceedings of the 23rd International Conference on Supercomputing},
pages = {306–315},
numpages = {10},
location = {Yorktown Heights, NY, USA},
series = {ICS '09}
}

@inproceedings{sharma2022compute,
  title={Compute Express Link{\textregistered}: An open industry-standard interconnect enabling heterogeneous data-centric computing},
  author={Sharma, Debendra Das},
  booktitle={2022 IEEE Symposium on High-Performance Interconnects (HOTI)},
  pages={5--12},
  year={2022},
  organization={IEEE}
}

@inproceedings{van2019hoti,
  title={Hoti 2019: Compute express link},
  author={Van Doren, Stephen},
  booktitle={2019 IEEE Symposium on High-Performance Interconnects (HOTI)},
  pages={18--18},
  year={2019},
  organization={IEEE}
}

@inproceedings{wang2020characterizing,
  title={Characterizing and modeling non-volatile memory systems},
  author={Wang, Zixuan and Liu, Xiao and Yang, Jian and Michailidis, Theodore and Swanson, Steven and Zhao, Jishen},
  booktitle={2020 53rd Annual IEEE/ACM International Symposium on Microarchitecture (MICRO)},
  pages={496--508},
  year={2020},
  organization={IEEE}
}

@inproceedings{xiang2022characterizing,
  title={{Characterizing the Performance of Intel Optane Persistent Memory: A Close Look at its On-DIMM Buffering}},
  author={Xiang, Lingfeng and Zhao, Xingsheng and Rao, Jia and Jiang, Song and Jiang, Hong},
  booktitle={Proceedings of the Seventeenth European Conference on Computer Systems},
  pages={488--505},
  year={2022}
}

@inproceedings{bergman2022reconsidering,
  title={Reconsidering os memory optimizations in the presence of disaggregated memory},
  author={Bergman, Shai and Faldu, Priyank and Grot, Boris and Vilanova, Llu{\'\i}s and Silberstein, Mark},
  booktitle={Proceedings of the 2022 ACM SIGPLAN International Symposium on Memory Management},
  pages={1--14},
  year={2022}
}

@inproceedings{dragojevic2014farm,
  title={$\{$FaRM$\}$: Fast remote memory},
  author={Dragojevi{\'c}, Aleksandar and Narayanan, Dushyanth and Castro, Miguel and Hodson, Orion},
  booktitle={11th USENIX Symposium on Networked Systems Design and Implementation (NSDI 14)},
  pages={401--414},
  year={2014}
}

@inproceedings{gu2017efficient,
  title={Efficient memory disaggregation with infiniswap},
  author={Gu, Juncheng and Lee, Youngmoon and Zhang, Yiwen and Chowdhury, Mosharaf and Shin, Kang G},
  booktitle={14th USENIX Symposium on Networked Systems Design and Implementation (NSDI 17)},
  pages={649--667},
  year={2017}
}

@inproceedings{abulila2019flatflash,
  title={Flatflash: Exploiting the byte-accessibility of ssds within a unified memory-storage hierarchy},
  author={Abulila, Ahmed and Mailthody, Vikram Sharma and Qureshi, Zaid and Huang, Jian and Kim, Nam Sung and Xiong, Jinjun and Hwu, Wen-mei},
  booktitle={Proceedings of the Twenty-Fourth International Conference on Architectural Support for Programming Languages and Operating Systems},
  pages={971--985},
  year={2019}
}

@article{Maruf2022MULTICLOCKDT,
  title={MULTI-CLOCK: Dynamic Tiering for Hybrid Memory Systems},
  author={Adnan Maruf and Ashikee Ghosh and Janki Bhimani and Daniela Campello and Andy Rudoff and Raju Rangaswami},
  journal={2022 IEEE International Symposium on High-Performance Computer Architecture (HPCA)},
  year={2022},
  pages={925-937},
  url={https://api.semanticscholar.org/CorpusID:248865268}
}

@article{Kumar2021RadiantEP,
  title={Radiant: efficient page table management for tiered memory systems},
  author={Sandeep Kumar and Aravinda Prasad and Smruti Ranjan Sarangi and Sreenivas Subramoney},
  journal={Proceedings of the 2021 ACM SIGPLAN International Symposium on Memory Management},
  year={2021},
  url={https://api.semanticscholar.org/CorpusID:235463147}
}

@article{Hildebrand2020AutoTMAT,
  title={AutoTM: Automatic Tensor Movement in Heterogeneous Memory Systems using Integer Linear Programming},
  author={Mark Hildebrand and Jawad Ali Khan and Sanjeev N. Trika and Jason Lowe-Power and Venkatesh Akella},
  journal={Proceedings of the Twenty-Fifth International Conference on Architectural Support for Programming Languages and Operating Systems},
  year={2020},
  url={https://api.semanticscholar.org/CorpusID:212641763}
}

@article{Li2022TransparentAL,
  title={Transparent and lightweight object placement for managed workloads atop hybrid memories},
  author={Zhe Li and Mingyu Wu},
  journal={Proceedings of the 18th ACM SIGPLAN/SIGOPS International Conference on Virtual Execution Environments},
  year={2022},
  url={https://api.semanticscholar.org/CorpusID:247108266}
}

@article{Ren2021SentinelET,
  title={Sentinel: Efficient Tensor Migration and Allocation on Heterogeneous Memory Systems for Deep Learning},
  author={Jie Ren and Jiaolin Luo and Kai Wu and Minjia Zhang and Hyeran Jeon},
  journal={2021 IEEE International Symposium on High-Performance Computer Architecture (HPCA)},
  year={2021},
  pages={598-611},
  url={https://api.semanticscholar.org/CorpusID:231620477}
}

@article{Wang2019PantheraHM,
  title={Panthera: holistic memory management for big data processing over hybrid memories},
  author={Chenxi Wang and Huimin Cui and Ting Cao and John N. Zigman and Haris Volos and Onur Mutlu and Fang Lv and Xiaobing Feng and Guoqing Harry Xu},
  journal={Proceedings of the 40th ACM SIGPLAN Conference on Programming Language Design and Implementation},
  year={2019},
  url={https://api.semanticscholar.org/CorpusID:150372592}
}

@article{Dulloor2016DataTI,
  title={Data tiering in heterogeneous memory systems},
  author={Subramanya R. Dulloor and Amitabha Roy and Zheguang Zhao and Narayanan Sundaram and Nadathur Satish and Rajesh Sankaran and Jeffrey R. Jackson and Karsten Schwan},
  journal={Proceedings of the Eleventh European Conference on Computer Systems},
  year={2016},
  url={https://api.semanticscholar.org/CorpusID:8681081}
}

@article{Raybuck2021HeMemST,
  title={{HeMem: Scalable Tiered Memory Management for Big Data Applications and Real NVM}},
  author={Amanda Raybuck and Tim Stamler and Wei Zhang and Mattan Erez and Simon Peter},
  journal={Proceedings of the ACM SIGOPS 28th Symposium on Operating Systems Principles},
  year={2021},
  url={https://api.semanticscholar.org/CorpusID:239029009}
}

@article{Kannan2017HeteroOSO,
  title={HeteroOS — OS design for heterogeneous memory management in datacenter},
  author={Sudarsun Kannan and Ada Gavrilovska and Vishal Gupta and Karsten Schwan},
  journal={2017 ACM/IEEE 44th Annual International Symposium on Computer Architecture (ISCA)},
  year={2017},
  pages={521-534},
  url={https://api.semanticscholar.org/CorpusID:19189083}
}

@inproceedings{Kim2021ExploringTD,
  title={Exploring the Design Space of Page Management for Multi-Tiered Memory Systems},
  author={Jonghyeon Kim and Wonkyo Choe and Jeongseob Ahn},
  booktitle={USENIX Annual Technical Conference},
  year={2021},
  url={https://api.semanticscholar.org/CorpusID:236992513}
}

@article{Weiner2022TMOTM,
  title={{TMO: transparent memory offloading in datacenters}},
  author={Johannes Weiner and Niket Agarwal and Dan Schatzberg and Leon Yang and Hao Wang and Blaise Sanouillet and Bikash Sharma and Tejun Heo and M. Jain and Chunqiang Tang and Dimitrios Skarlatos},
  journal={Proceedings of the 27th ACM International Conference on Architectural Support for Programming Languages and Operating Systems},
  year={2022},
  url={https://api.semanticscholar.org/CorpusID:247026540}
}

@string{OR="{O'Reilly \& Associates, Inc.\ }"}

@manual{MPI,
    organization = "Message Passing Interface Forum", 
    title = "MPI: A Message-Passing Interface Standard", 
    month = "Mar", 
    year  = "1994"}

@misc{RDMA,
	author = "R. Recio and P. Culley and D. Garcia and J. Hilland",
	title = "An RDMA Protocol Specification (Version 1.0)",
	month = "October",
	year = "2002",
    }

@InProceedings{gpu_dp_micro14,
    author =  "Guoyang Chen and Bo Wu and Dong Li and Xipeng Shen",
    title = {{PORPLE: An Extensible Optimizer for Portable Data Placement on GPU}},
    booktitle =  "IEEE/ACM International Symposium on Microarchitecture",
    year =   "2014"
}

@inproceedings{hpdc16:wu,
 author = {Panruo Wu and Dong Li and Zizhong Chen and Jeffrey Vetter and Sparsh Mittal},
 title = {Algorithm-Directed Data Placement in Explicitly Managed Non-Volatile Memory},
 booktitle = {ACM Symposium on High-Performance Parallel and Distributed Computing (HPDC)},
 year = {2016}
}

@INPROCEEDINGS{8048928,
author={Y. {Huang} and D. {Li}},
booktitle={IEEE International Conference on Cluster Computing (CLUSTER)},
title={{Performance Modeling for Optimal Data Placement on GPU with Heterogeneous Memory Systems}},
year={2017}
}

@inproceedings{sc18:wu,
    author =       "Kai Wu and Jie Ren and Dong Li",
    title =        {{Runtime Data Management on Non-Volatile Memory-Based Heterogeneous Memory for Task Parallel Programs}},
    booktitle =  "ACM/IEEE International Conference for High Performance Computing, Networking, Storage and Analysis",
    year =         2018
}

@article{doi:10.1177/109434209100500306,
author = {D.H. Bailey and E. Barszcz and J.T. Barton and D.S. Browning and R.L. Carter and L. Dagum and R.A. Fatoohi and P.O. Frederickson and T.A. Lasinski and R.S. Schreiber and H.D. Simon and V. Venkatakrishnan and S.K. Weeratunga},
title ={The Nas Parallel Benchmarks},
journal = {The International Journal of Supercomputing Applications},
year = {1991},
}

@inproceedings{unimem:sc17,
  author    = {K. Wu and Y. Huang and D. Li},
  title     = {{Unimem: Runtime Data Management on Non-Volatile Memory-based Heterogeneous Main Memory}},
  booktitle = {International Conference for High Performance Computing, Networking, Storage and Analysis},
  year      = {2017}
}

@inproceedings{zuo2018write,
  title={Write-optimized and high-performance hashing index scheme for persistent memory},
  author={Zuo, Pengfei and Hua, Yu and Wu, Jie},
  booktitle={13th $\{$USENIX$\}$ Symposium on Operating Systems Design and Implementation ($\{$OSDI$\}$ 18)},
  pages={461--476},
  year={2018}
}

@misc{mlc,
	author        = "{Intel Corporation}",
	title         = "Intel Memory Latency Checker v3.5",
	year          = "2019",
	url           = "https://software.intel.com/en-us/articles/intelr-memory-latency-checker"
}

@inproceedings{micro18:pim,
  author    = {Jiawen Liu and Hengyu Zhao and Matheus A. Ogleari and Dong Li and Jishen Zhao},
  title     = {{Processing-in-Memory for Energy-Efficient Neural Network Training: {A} Heterogeneous Approach}},
  booktitle = {IEEE/ACM International Symposium on Microarchitecture},
  year      = {2018}
}

@inproceedings{luo:NGS,
    title = {{Enabling Faster NGS Analysis on Optane-based Heterogeneous Memory}},
author = {Luo, Jiaolin and Guo, Luanzheng and Ren, Jie and Wu, Kai and Li, Dong},
booktitle = "Proceedings of Supercomputing '20 (Posters)",
month = Nov,
year = 2020
}

@inproceedings{neurips20:hm-ann,
 author = {Jie Ren and Minjia Zhang and Dong Li},
 title = {{HM-ANN: Efficient Billion-Point Nearest Neighbor Search on Heterogeneous Memory}},
 booktitle = {Conference on Neural Information Processing Systems (NeurIPS)},
 year = {2020}
}

@inproceedings{ppopp21:sparta,
    author          = "Jiawen Liu and Jie Ren and Roberto Gioiosa and Dong Li and Jiajia Li",
    title           = {{Sparta: High-Performance, Element-Wise Sparse Tensor Contraction on Heterogeneous Memory}},
    booktitle       = "{Principles and Practice of Parallel Programming}",
    year            = "2021"
}

@inproceedings{ics21:athena,
    title={{Athena: High-Performance Sparse Tensor Contraction Sequence on Heterogeneous Memory}},
    author={Jiawen Liu and Dong Li and Jiajia Li},
    booktitle={{International Conference on Supercomputing (ICS)}},
    year={2021}
}

@inproceedings{ics21:warpx,
    title={{Optimizing Large-Scale Plasma Simulations on Persistent Memory-based Heterogeneous Memory with Effective Data Placement Across Memory Hierarchy}},
    author={Jie Ren and Jiaolin Luo and Ivy Peng and Kai Wu and Dong Li},
    booktitle={{International Conference on Supercomputing (ICS)}},
    year={2021}
}

@misc{cxl,
    title = "{Compute Express Link Industry Members}",
    howpublished = "https://www.computeexpresslink.org/members",
    year = {2021}
}

@inproceedings{ppopp23:merchandiser,
    author          = "Zhen Xie and Jie Liu and Jiajia Li and Dong Li",
    title           = {{Merchandiser: Data Placement on Heterogeneous Memory for Task-Parallel HPC Applications with Load-Balance Awareness}},
    booktitle       = {{Proceedings of the Symposium on Principles and Practices of Parallel Programming (PPoPP)}},
    year            = {2023}
}

@inproceedings{betty:asplos23,
author = {Shuangyan Yang and Minjia Zhang and Wenqian Dong and Dong Li},
title = {{Betty: Enabling Large-Scale GNN Training with Batch-Level Graph Partitioning}},
year = {2023},
booktitle = {International Conference on Architectural Support for Programming Languages and Operating Systems (ASPLOS)}
}

@inproceedings{10.1145/3575693.3578835,
author = {Li, Huaicheng and Berger, Daniel S. and Hsu, Lisa and Ernst, Daniel and Zardoshti, Pantea and Novakovic, Stanko and Shah, Monish and Rajadnya, Samir and Lee, Scott and Agarwal, Ishwar and Hill, Mark D. and Fontoura, Marcus and Bianchini, Ricardo},
title = {{Pond: CXL-Based Memory Pooling Systems for Cloud Platforms}},
year = {2023},
booktitle = {International Conference on Architectural Support for Programming Languages and Operating Systems (ASPLOS)}
}

@misc{sharma2023introduction,
      title={{An Introduction to the Compute Express Link (CXL) Interconnect}}, 
      author={Debendra Das Sharma and Robert Blankenship and Daniel S. Berger},
      year={2023},
      eprint={2306.11227},
      archivePrefix={arXiv},
      primaryClass={cs.AR}
}

@inproceedings{eurosys24:mtm,
    author = "Jie Ren and Dong Xu and Junhee Ryu and Kwangsik Shin and Daewoo Kim and Dong Li",
    title = {{MTM: Rethinking Memory Profiling and Migration for Multi-Tiered Large Memory Systems}}, 
    booktitle = "European Conference on Computer Systems",
    year = "2024"
}

@inproceedings{atc24_hm,
author = {Dong Xu and Junhee Ryu and Jinho Baek and Kwangsik Shin and Pengfei Su and Dong Li},
title = {{FlexMem: Adaptive Page Profiling and Migration for Tiered Memory}},
year = {2024},
booktitle = {30th USENIX Annual Technical Conference (ATC)}
}

@inproceedings{hpca25:buffalo,
    author = {Shuangyan Yang and Minjia Zhang and Dong Li},
    title = {{Buffalo: Enabling Large-Scale GNN Training via Memory-Efficient Bucketization}},
    booktitle = {International Symposium on High-Performance Computer Architecture (HPCA)},
    year = 2025
}

@INPROCEEDINGS{memsys_cxl_switch_2024,
  author={Jianbo Wu and Jie Liu and Gokcen Kestor and Roberto Gioiosa and Dong Li},
  booktitle={10th International Symposium on Memory Systems}, 
  title={{Performance Study of CXL Memory Topology}}, 
  year={2024}
}

@InProceedings{hpca25:dlrm,
  author = {Bin Ma and Jie Ren and Shuangyan Yang and Benjamin Francis and Ehsan Ardestani and Min Si and Dong Li},
  title = {{Machine Learning-Guided Memory Optimization for DLRM Inference on Tiered Memory}},
  booktitle={International Symposium on High Performance Computer Architecture (HPCA)},
  year = 2025
}

@inproceedings{10.1145/3488423.3519336,
author = {Peng, Ivy and Karlin, Ian and Gokhale, Maya and Shoga, Kathleen and Legendre, Matthew and Gamblin, Todd},
title = {{A Holistic View of Memory Utilization on HPC Systems: Current and Future Trends}},
year = {2022},
booktitle = {Proceedings of the International Symposium on Memory Systems}
}

@inproceedings{10.1145/3581784.3607108,
author = {Wahlgren, Jacob and Schieffer, Gabin and Gokhale, Maya and Peng, Ivy},
title = {{A Quantitative Approach for Adopting Disaggregated Memory in HPC Systems}},
year = {2023},
booktitle = {Proceedings of the International Conference for High Performance Computing, Networking, Storage and Analysis}
}

@inproceedings{10.1145/3624062.3624175,
author = {Fridman, Yehonatan and Mutalik Desai, Suprasad and Singh, Navneet and Willhalm, Thomas and Oren, Gal},
title = {{CXL Memory as Persistent Memory for Disaggregated HPC: A Practical Approach}},
year = {2023},
booktitle = {Proceedings of the SC Workshops of the International Conference on High Performance Computing, Network, Storage, and Analysis}
}

@InProceedings{10.1007/978-3-031-32041-5_16,
author="Li, Jie
and Michelogiannakis, George
and Cook, Brandon
and Cooray, Dulanya
and Chen, Yong",
editor="Bhatele, Abhinav
and Hammond, Jeff
and Baboulin, Marc
and Kruse, Carola",
title={{Analyzing Resource Utilization in an HPC System: A Case Study of NERSC's Perlmutter}},
booktitle="High Performance Computing",
year="2023",
publisher="Springer Nature Switzerland",
address="Cham",
pages="297--316",
isbn="978-3-031-32041-5"
}

@INPROCEEDINGS{10024061,
  author={Wahlgren, Jacob and Gokhale, Maya and Peng, Ivy B.},
  booktitle={IEEE/ACM Workshop on Memory Centric High Performance Computing (MCHPC)}, 
  title={{Evaluating Emerging CXL-enabled Memory Pooling for HPC Systems}}, 
  year={2022}
}

@INPROCEEDINGS{10898637,
  author={Giles, Ellis and Varman, Peter},
  booktitle={IEEE 31st International Conference on High Performance Computing, Data and Analytics Workshop (HiPCW)}, 
  title={{Towards Continuous Checkpointing for HPC Systems Using CXL}}, 
  year={2024}
}

@article{CASANOVA2025103125,
  title = {{Lowering Entry Barriers to Developing Custom Simulators of Distributed Applications and Platforms with SimGrid}},
  journal = {Parallel Computing},
  volume = {123},
  pages = {103-125},
  year = {2025},
  issn = {0167-8191},
  author = {Casanova, Henri and Giersch, Arnaud and Legrand, Arnaud and Quinson, Martin and Suter, Fr{\'e}d{\'e}ric}
}

@InProceedings{10.1007/978-3-031-73370-3_6,
author="Adam, Julien
and Besnard, Jean-Baptiste
and Roussel, Adrien
and Jaeger, Julien
and Carribault, Patrick
and P{\'e}rache, Marc",
editor="Blaas-Schenner, Claudia
and Niethammer, Christoph
and Haas, Tobias",
title={{To Share or Not to Share: A Case for MPI in Shared-Memory}},
booktitle="Recent Advances in the Message Passing Interface",
year="2025"
}

@INPROCEEDINGS{10825804,
  author={Ahn, Hooyoung and Kim, Seonyoung and Park, Yoomi and Han, Woojong and Ahn, Shinyoung and Tran, Tu and Ramesh, Bharath and Subramoni, Hari and Panda, Dhabaleswar K.},
  booktitle={2024 IEEE International Conference on Big Data (BigData)}, 
  title={{MPI Allgather Utilizing CXL Shared Memory Pool in Multi-Node Computing Systems}}, 
  year={2024}
}

@INPROCEEDINGS{10898940,
  author={Walia, Pranjal and Shanware, Ishan and Vaidyanathan, Karthikeyan and Kalamkar, Dhiraj D. and Natarajan, Uma M.},
  booktitle={International Conference on High Performance Computing, Data and Analytics Workshop (HiPCW)}, 
  title={{LOSM: Leveraging OpenMP and Shared Memory for Accelerating Blocking MPI Allreduce}}, 
  year={2024}
}

@inproceedings{10.1145/2462902.2462903,
author = {Li, Shigang and Hoefler, Torsten and Snir, Marc},
title = {{NUMA-Aware Shared-Memory Collective Communication for MPI}},
year = {2018},
booktitle = {International Symposium on High-Performance Parallel and Distributed Computing}
}

@inproceedings{10.1007/978-3-540-87475-1_21,
author = {Graham, Richard L. and Shipman, Galen},
title = {{MPI Support for Multi-core Architectures: Optimized Shared Memory Collectives}},
year = {2008},
publisher = {Springer-Verlag},
booktitle = {Proceedings of the 15th European PVM/MPI Users' Group Meeting on Recent Advances in Parallel Virtual Machine and Message Passing Interface}
}

@INPROCEEDINGS{4663761,
  author={Wei Huang and Koop, Matthew J. and Panda, Dhabaleswar K.},
  booktitle={International Conference on Cluster Computing}, 
  title={{Efficient One-Copy MPI Shared Memory Communication in Virtual Machines}}, 
  year={2008}
}

@article{white2020optimizing,
  title={{Optimizing Point-to-Point Communication between Adaptive MPI Endpoints in Shared Memory}},
  author={White, Sam and Kale, Laxmikant V},
  journal={Concurrency and Computation: Practice and Experience},
  volume={32},
  number={3},
  year={2020},
  publisher={Wiley Online Library}
}

@INPROCEEDINGS{10898426,
  author={Aravind, Ramesh and John, Groves},
  booktitle={International Conference on High Performance Computing, Data and Analytics Workshop (HiPCW)}, 
  title={{Study of CXL Memory Sharing with FamFS and its Use cases}}, 
  year={2024}}

@inproceedings{fan2022optimizing,
  title={{Optimizing the Bruck Algorithm for Non-Uniform All-to-All Communication}},
  author={Fan, Ke and Gilray, Thomas and Pascucci, Valerio and Huang, Xuan and Micinski, Kristopher and Kumar, Sidharth},
  booktitle={International Symposium on High-Performance Parallel and Distributed Computing},
  year={2022}
}

@inproceedings{arap2015adaptive,
  title={{Adaptive Recursive Doubling Algorithm for Collective Communication}},
  author={Arap, Omer and Swany, Martin and Brown, Geoffrey and Himebaugh, Bryce},
  booktitle={International Parallel and Distributed Processing Symposium Workshop},
  year={2015}
}

@article{10.1145/3543176,
author = {Schelten, Niklas and Steinert, Fritjof and Knapheide, Justin and Schulte, Anton and Stabernack, Benno},
title = {A High-Throughput, Resource-Efficient Implementation of the RoCEv2 Remote DMA Protocol and its Application},
year = {2022},
journal = {ACM Transactions on Reconfigurable Technologies and Systems},
articleno = {5}
}

@inproceedings{ahn2022enabling,
  title={{Enabling CXL Memory Expansion for In-Memory Database Management Systems}},
  author={Ahn, Minseon and Chang, Andrew and Lee, Donghun and Gim, Jongmin and Kim, Jungmin and Jung, Jaemin and Rebholz, Oliver and Pham, Vincent and Malladi, Krishna and Ki, Yang Seok},
  booktitle={International Workshop on Data Management on New Hardware},
  year={2022}
}

@inproceedings{ma2024hydrarpc,
  title={$\{$HydraRPC$\}$:$\{$RPC$\}$ in the $\{$CXL$\}$ Era},
  author={Ma, Teng and Liu, Zheng and Wei, Chengkun and Huang, Jialiang and Zhuo, Youwei and Li, Haoyu and Zhang, Ning and Guan, Yijin and Niu, Dimin and Zhang, Mingxing and others},
  booktitle={USENIX Annual Technical Conference},
  year={2024}
}

@misc{perftest,
 author ={Open Fabrics Enterprise Distribution},
 title={perftest},
 note={\url{https://github.com/linux-rdma/perftest}}
}

@inproceedings{broder2001using,
  title={Using multiple hash functions to improve IP lookups},
  author={Broder, Andrei and Mitzenmacher, Michael},
  booktitle={Proceedings IEEE INFOCOM 2001. Conference on Computer Communications. Twentieth Annual Joint Conference of the IEEE Computer and Communications Society (Cat. No. 01CH37213)},
  volume={3},
  pages={1454--1463},
  year={2001},
  organization={IEEE}
}

@article{jain2024memory,
  title={Memory Sharing with CXL: Hardware and Software Design Approaches},
  author={Jain, Sunita and Yeleswarapu, Nagaradhesh and Maruf, Hasan Al and Gupta, Rita},
  journal={arXiv preprint arXiv:2404.03245},
  year={2024}
}

@article{guide2011intel,
  title={Intel{\textregistered} 64 and ia-32 architectures software developer’s manual},
  author={Guide, Part},
  journal={Volume 3B: system programming guide, Part},
  volume={2},
  number={11},
  pages={0--40},
  year={2011}
}

@inproceedings{wang2025cxl_performance,
  title={Performance Characterization of CXL Memory and Its Use Cases},
  author={Wang, Xi and Liu, Jie and Wu, Jianbo and Yang, Shuangyan and Ren, Jie and Shankar, Bhanu and Li, Dong},
  booktitle={2025 IEEE International Parallel and Distributed Processing Symposium (IPDPS)},
  pages={1048--1061},
  year={2025},
  organization={IEEE}
}

@article{zuo2019level,
  title={Level hashing: A high-performance and flexible-resizing persistent hashing index structure},
  author={Zuo, Pengfei and Hua, Yu and Wu, Jie},
  journal={ACM Transactions on Storage (TOS)},
  volume={15},
  number={2},
  pages={1--30},
  year={2019},
  publisher={ACM New York, NY, USA}
}

@inproceedings{broder1990multilevel,
  title={Multilevel adaptive hashing},
  author={Broder, Andrei Z and Karlin, Anna R},
  booktitle={Proceedings of the first annual ACM-SIAM symposium on Discrete algorithms},
  pages={43--53},
  year={1990}
}

@inproceedings{adaptiveMPI,
  title={Adaptive mpi},
  author={Huang, Chao and Lawlor, Orion and Kale, Laxmikant V},
  booktitle={International workshop on languages and compilers for parallel computing},
  pages={306--322},
  year={2003},
  organization={Springer}
}

@inproceedings{chen2020lock,
  title={Lock-free concurrent level hashing for persistent memory},
  author={Chen, Zhangyu and Hua, Yu and Ding, Bo and Zuo, Pengfei},
  booktitle={2020 USENIX Annual Technical Conference (USENIX ATC 20)},
  pages={799--812},
  year={2020}
}

@article{xiong2023dalea,
  title={Dalea: A persistent multi-level extendible hashing with improved tail performance},
  author={Xiong, Zi-Wei and Jiang, De-Jun and Xiong, Jin and Ren, Ren},
  journal={Journal of Computer Science and Technology},
  volume={38},
  number={5},
  pages={1051--1073},
  year={2023},
  publisher={Springer}
}

@misc{osu_omb,
  title = {OSU Micro-Benchmarks.},
author = {Network-Based Computing Laboratory},
  url = "https://mvapich.cse.ohio-state.edu/benchmarks/",
month = {},
year = {}
}

@misc{MellanoxCX3,
    author = {Mellanox Technologies},
  title = {HP and Mellanox Benchmarking Report for Ultra Low Latency 10 and 40Gb/s Ethernet Interconnect.},
  url = "https://network.nvidia.com/related-docs/whitepapers/HP_Mellanox_FSI%20Benchmarking%20Report%20for%2010%20%26%2040GbE.pdf",
month = {},
year = {}
}

@misc{MellanoxCX6,
  title = {NVIDIA ConnectX-6 InfiniBand/Ethernet Adapter Cards User Manual},
author = {NVIDIA},
  url = "https://docs.nvidia.com/networking/display/connectx6vpi/introduction",
month = {},
year = {}
}

@misc{Niagara20,
  title = {Exploring CXL Memory Disaggregation:
Use Cases and System Benefits},
author = {Choi, Jungmin},
  url = "https://memverge.com/wp-content/uploads/Memory-Fabric-Forum-at-OCP-Global-Summit-2024-%E2%80%93-SK-hynix_Exploring-CXL-Memory-Disaggregation.pdf",
month = {},
year = {}
}

@misc{MPICH_online,
  title = {MPICH | High-Performance Portable MPI},
author = {Argonne National Lab},
  url = "https://www.mpich.org/",
month = {},
year = {}
}

@misc{iPerf_online,
  title = {iPerf - The TCP, UDP and SCTP network bandwidth measurement tool},
author = {Lawrence Berkeley National Laboratory},
  url = "https://iperf.fr/",
    year = {2025}
}

@inproceedings{wahlgren2023quantitative,
  title={A quantitative approach for adopting disaggregated memory in HPC systems},
  author={Wahlgren, Jacob and Schieffer, Gabin and Gokhale, Maya and Peng, Ivy},
  booktitle={Proceedings of the International Conference for High Performance Computing, Networking, Storage and Analysis},
  pages={1--14},
  year={2023}
}

\end{document}